\documentclass[10pt]{iopart}
\usepackage{graphicx}
\usepackage{iopams}  
\usepackage{braket}

\usepackage{xcolor}

\usepackage[colorlinks=true, urlcolor=blue, citecolor=black, linkcolor = black]{hyperref}

\begin{document}

\title{Coherent multidimensional spectroscopy in the gas phase}

\author{Lukas Bruder$^1$, Ulrich Bangert$^1$, Marcel Binz$^1$, Daniel Uhl$^1$ and Frank Stienkemeier$^{1,2}$}

\address{$^1$Institute of Physics, University of Freiburg, Hermann-Herder-Str. 3, 79104 Freiburg, Germany}

\address{$^2$Freiburg Institute of Advanced Studies (FRIAS), University of Freiburg, Albertstr. 19, D-79194 Freiburg, Germany}

\ead{lukas.bruder@physik.uni-freiburg.de}

\vspace{10pt}
\begin{indented}
\item[]April 2019
\end{indented}

\begin{abstract}
Recent work applying multidimentional coherent electronic spectroscopy at dilute samples in the gas phase is reviewed. The development of refined phase cycling approaches with improved sensitivity has opened up new opportunities to probe even dilute gas-phase samples. In this context, first results of two-dimensional spectroscopy performed at doped helium droplets reveal the femtosecond dynamics upon electronic excitation of cold, weakly-bound molecules, and even the induced dynamics from the interaction with the helium environment. Such experiments, offering well-defined conditions at low temperatures, are potentially enabling the isolation of fundamental processes in the excitation and charge transfer dynamics of molecular structures which so far have been masked in complex bulk environments. 
\end{abstract}

%
\vspace{2pc}

%
\submitto{\JPB}
%
%
%

\section{Introduction}

The development of coherent multidimensional spectroscopy (CMDS) in the optical regime has greatly improved the toolkit of ultrafast spectroscopy\,\cite{hochstrasser_two-dimensional_2007, cho_coherent_2008, nuernberger_multidimensional_2015}. The method may be regarded as an extension of pump-probe spectroscopy, where pump and probe steps are both spectrally resolved, while maintaining high temporal resolution in the sub 50\,fs regime\,\cite{jonas_two-dimensional_2003}. By spreading the nonlinear response onto multidimensional frequency-correlation maps, improved spectral decongestion is achieved and the analysis of couplings within the system or to the environment is greatly simplified. 

The concept of CMDS was originally developed in NMR spectroscopy\,\cite{aue_twodimensional_1976} and was first implemented at optical frequencies about 25 years ago\,\cite{hamm_structure_1998, hybl_two-dimensional_1998}. Nowadays, routinely used methods in the optical regime comprise two-dimensional infrared (2DIR) spectroscopy and its counter part in the visible spectral range, 2D electronic spectroscopy (2DES)\,\cite{hochstrasser_two-dimensional_2007}. 2DIR yields insights into structure and dynamics of molecular networks with high chemical sensitivity, while 2DES accesses in addition electronic degrees of freedom (DOFs) and provides information about the coupled electronic-nuclear dynamics albeit with reduced chemical selectivity. Notably, also combinations of both methods have been reported\,\cite{oliver_correlating_2014, vanwilderen_mixed_2014, courtney_measuring_2015} and extensions to the THz spectral regime exist\,\cite{woerner_ultrafast_2013}. 

In recent years, 2DIR and 2DES have provided decisive information about many ultrafast phenomena including energy and charge transfer in biological systems\,\cite{brixner_two-dimensional_2005, read_cross-peak-specific_2007, lewis_probing_2012, romero_quantum_2014, fuller_vibronic_2014, dostal_situ_2016, scholes_using_2017, thyrhaug_identification_2018}, the real-time analysis of solvent dynamics\,\cite{asbury_water_2004, cowan_ultrafast_2005, moca_two-dimensional_2015} or the mapping of reaction pathways in chemical reactions\,\cite{nuernberger_multidimensional_2015}, to name a few examples. 
Furthermore, multiple-quantum coherence CMDS studies have proven to be sensitive probes for many-body effects\,\cite{stone_two-quantum_2009, turner_coherent_2010, cundiff_optical_2012, dai_two-dimensional_2012}. 

So far, CMDS has been mostly performed in the condensed phase, where the wide range of accessible systems has lead to a rich variety of studies, ranging from molecular aggregates\,\cite{milota_two-dimensional_2009, ginsberg_two-dimensional_2009, nemeth_tracing_2009, dostal_direct_2018} to multichromophoric photosynthic systems\,\cite{brixner_two-dimensional_2005, read_cross-peak-specific_2007, myers_two-dimensional_2010, collini_coherently_2010, dostal_situ_2016}, thin films\,\cite{provencher_direct_2014, oudenhoven_dye_2015, de_sio_tracking_2016, ostrander_energy_2017} and bulk crystal structures\,\cite{huxter_vibrational_2013, bakulin_real-time_2015, batignani_probing_2018} as well as different types of semiconductor nanomaterials\,\cite{moody_exciton-exciton_2011, bylsma_quantum_2012, cundiff_optical_2012, graham_two-dimensional_2012, mehlenbacher_energy_2015}.

Yet, most of these systems exhibit a large number of degrees of freedom (DOFs) with commonly complex interrelations, resulting often in highly congested data and significant spectral broadening. As such, detailed theoretical calculations are essential for the interpretation of the experiments. However, the large parameter spaces of condensed phase systems force to strong approximations and simplifications in theoretical models, even with nowadays numerical capacities. This makes the precise analysis and interpretation of the experiments, despite the advanced spectroscopic methods, up to date a challenging task.  

CMDS studies of isolated gas-phase systems resolve these issues and thus provide an invaluable complementary view on condensed phase experiments. In the gas phase, isolated model systems of confined complexity can be synthesized with a high degree of control\,\cite{hertel_ultrafast_2006}. Systems may be prepared in well-defined initial states\,\cite{scoles_atomic_1988, stienkemeier_use_1995, toennies_superfluid_2004}, and high resolution data can be acquired\,\cite{smalley_molecular_1977, levy_laser_1980, wewer_laser-induced_2004} while perturbations by the environment are eliminated. This also improves the situation for theorists and will assist the development of more accurate models, ultimately leading to a better understanding of primary ultrafast processes. 

In addition, the information content deducible from these experiments is increased by a palette of highly selective detection methods exclusively accessible in the gas phase. These include ion-mass\,\cite{wiley_timeflight_1955, schultz_efficient_2004, lippert_femtosecond_2004, bruder_phase-modulated_2015} and photoelectron kinetic energy spectrometry\,\cite{kruit_magnetic_1983, stolow_femtosecond_2004}, velocity map imaging (VMI)\,\cite{eppink_velocity_1997, vallance_molecular_2004, r.ashfold_imaging_2006, greaves_velocity_2010} and even electron-ion coincidence detection\,\cite{helm_images_1993, radloff_internal_1997, ergler_time-resolved_2005}. The additional information gained from these detection types can be extremely helpful in disentangling complex dynamics and sensing the system's reaction energy landscape including dark states\,\cite{stolow_femtosecond_2004}, non-radiative internal conversion pathways\cite{wolf_probing_2017} and reaction intermediates \,\cite{dantus_realtime_1987}. As such, the gas-phase approach opens new possibilities for high precision multidimensional spectroscopy studies unveiling an unprecedented amount of details. 

However, this development has been so far precluded by insufficient sensitivity to probe highly dilute gas-phase samples. 
Only few examples of CMDS in the gas phase exist to date. 
\begin{figure}
	\centering
	\includegraphics[width=0.8\linewidth]{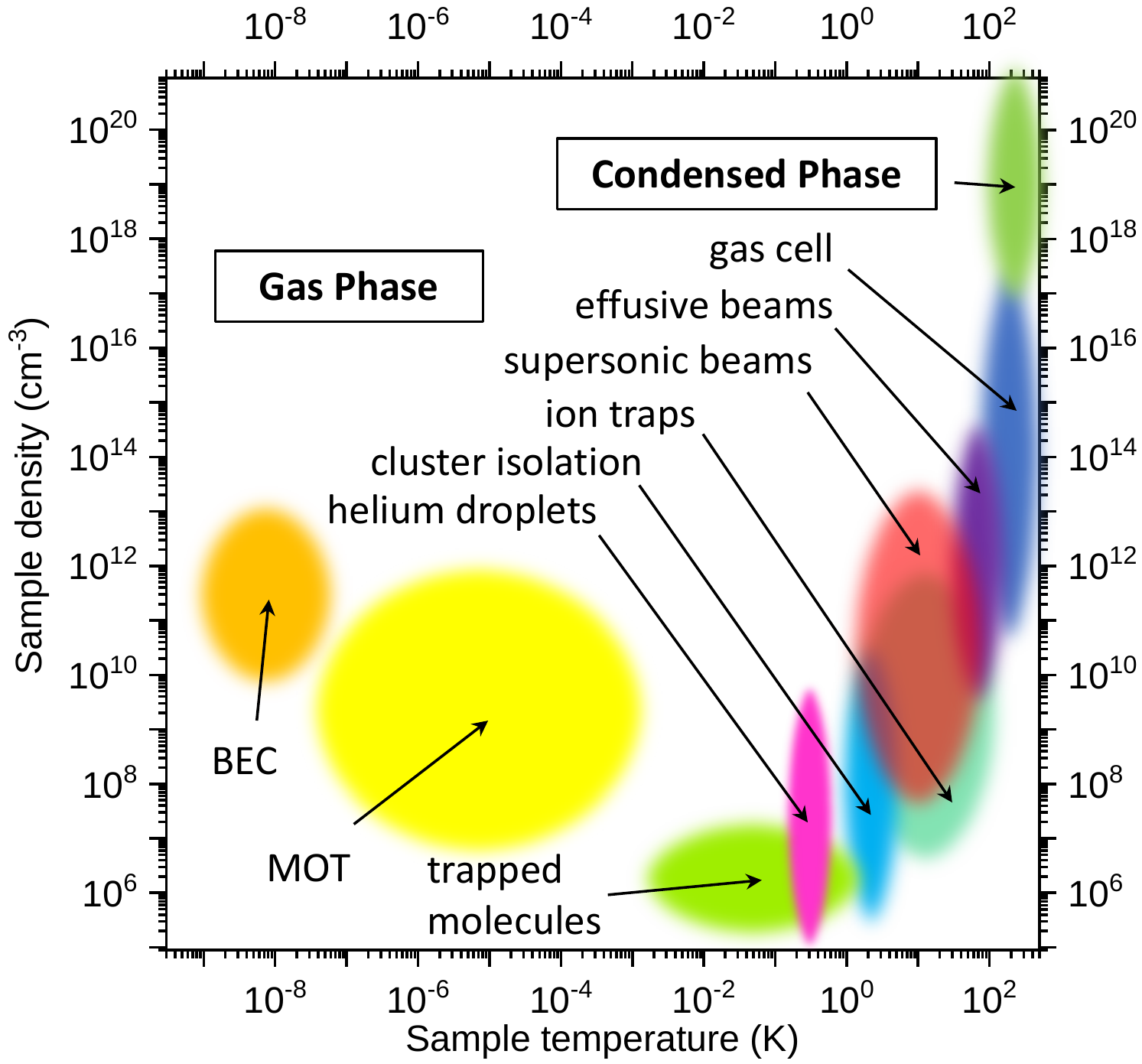}	
	\caption{Landscape showing typical temperatures and densities of gas phase samples in comparison to the condensed phase. So far, CMDS has been exclusively performed in the condensed phase and gas cells which cover only a small fraction of the available parameter space covered by nowadays available targets. The work of our group extends this range to CMDS experiments on cluster-isolation and helium droplet targets. MOT: atoms in magneto-optical traps; BEC: Bose-Einstein condensate of atoms.}
	\label{fig:overview}
\end{figure}
Some groups have reported 2DES studies of alkali atom vapors at particle densities of $\geq 10^{10}$\,cm$^{-3}$\,\cite{tian_femtosecond_2003, tekavec_fluorescence-detected_2007, dai_two-dimensional_2010} and very recently down to $10^8$\,cm$^{-3}$\,\cite{Yu_2019}.
These simple target systems, however, do not imply a generalization of the method's applicability to more advanced systems. 
For rubidium vapors, also a high-resolution 2D spectroscopy scheme based on frequency combs has been recently demonstrated capable of even resolving the atomic hyperfine levels\,\cite{lomsadze_frequency_2017, lomsadze_tri-comb_2018}. 
Furthermore, high-resolution multidimensional spectroscopy in the frequency domain using nanosecond lasers has been performed on several molecular vapors\,\cite{Chen_review_2016}. 
Yet, considering the wide range of unique target systems available in the gas phase and the large parameter space they cover, CMDS has been so far restricted to a small portion of targets. 
This is illustrated in Fig.\ \ref{fig:overview} where the landscape of gas-phase samples provided by different experimental techniques is plotted with respect to sample density and internal temperature. 

Only very recently, the Brixner group and our group demonstrated the first 2DES studies of gas-phase molecules and incorporated some of the afore mentioned new photoionization detection schemes\,\cite{roeding_coherent_2018, bruder_coherent_2018}. 
Brixner and coworkers probed a thermal gas of NO$_2$ molecules combined with selective mass spectrometry\,\cite{roeding_coherent_2018}. Our group studied cold ($T =380$\,mK) Rb$_2$ and Rb$_3$ molecules prepared with matrix isolation in a cluster beam apparatus, detected with photoelectron and ion-mass spectrometry\,\cite{bruder_coherent_2018}. 
These experiments, in principle, continue the pioneering early work of Zewail and coworkers\,\cite{zewail_femtochemistry:_2000}, advancing the field of Femtochemistry to a new direction and extending the range of target systems towards more fundamental quantum systems which may now come within reach (Fig.\,\ref{fig:overview}).

Besides the extreme demands on sensitivity, another circumstance has prolonged the development of gas-phase multidimensional spectroscopy. The vacuum technologies required for advanced sample preparation in the gas phase are not common to the CMDS community. On the other side, the molecular beam and related communities are mostly not aware of CMDS techniques. Hence, this combination of techniques requires expertise from rather disjunct communities and calls for a novel fusion of disciplines and their specialized technologies. 

In this context, we review in the present paper the recent extension of CMDS to the gas phase in conjunction with the 2DES experiments developed and conducted in our group. Along this, we will provide a brief introduction to both methodologies, CMDS on the one hand, gas-phase sample preparation and detection on the other hand. We examine the technical challenges and solutions of gas-phase 2DES, and conclude with a discussion of future perspectives. 

\section{Principle of 2DES}
\subsection{Principle and advantages of 2DES}
\begin{figure}
\centering
  \includegraphics[width=0.6\linewidth]{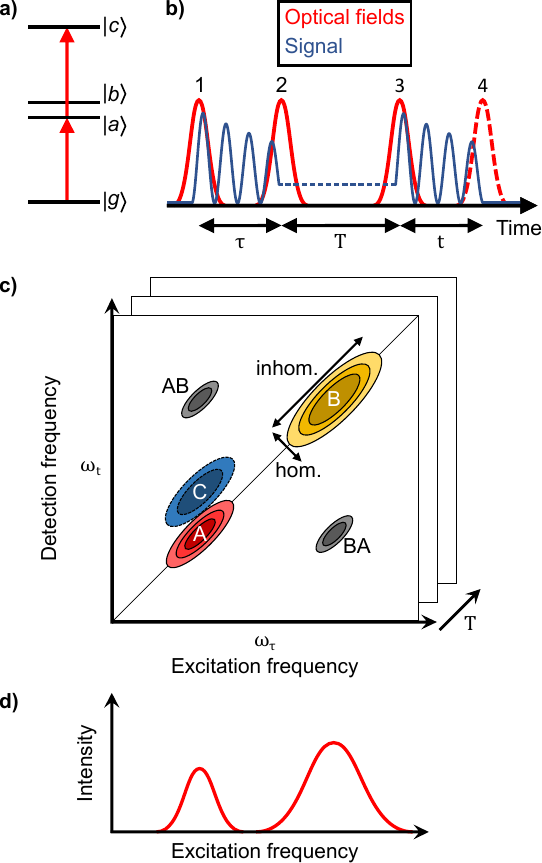}
  \caption{Principle of 2D spectroscopy. 
  (a) Simplified four-level system. (b) Pulse sequence used in 2D spectroscopy. The sample is excited with 3 or 4 pulses (indicated by dashed pulse envelope). The blue trace indicates the signal, which represents an oscillating dipole during the coherence times $\tau, t$. In between pulse 2 and 3 (time interval $T$), the system's time evolution (dashed trace) is probed. All three time variables $\tau, t, T$ are systematically scanned in the experiment. (c) 2D frequency-correlation map obtained by a 2D Fourier transform of the data set with respect to the coherence times $\tau, t$ directly correlating excitation ($\omega_\tau$) and detection ($\omega_t$) frequencies. Peaks A, B on the diagonal represent the $\ket{g} \leftrightarrow \ket{a}, \ket{b}$ resonances, with inhomogeneous and homogeneous lineshapes along the diagonal and antidiagonal, respectively. Peak C denotes an excited state absorption from $\ket{a}$ to the higher lying state  $\ket{c}$ (typically appearing with negative amplitude). AB and BA denote cross peaks, indicating couplings between states $\ket{a}$ and $\ket{b}$. The time evolution of all features is tracked as a function of $T$.(d) Linear absorption spectrum of the same system. Most spectral features overlap and are difficult to infer from the data. Likewise, a characterization of the system's inhomogeneity becomes difficult. 
  }
\label{fig:2D_scheme}
\end{figure}
The principle of CMDS is described in detail in several review articles and books\,\cite{jonas_two-dimensional_2003, cho_coherent_2008, ogilvie_chapter_2009, schlau-cohen_two-dimensional_2011, cundiff_optical_2012, branczyk_crossing_2014, nuernberger_multidimensional_2015, fuller_experimental_2015, moody_advances_2017, oliver_recent_2018, maiuri_electronic_2018, hamm_concepts_2011, fayer_watching_2011}. Here, we provide only a brief introduction to the basic concept of 2DES and highlight its most important features. 

In 2D spectroscopy, the sample (here approximated by a four-level model system, Fig.\,\ref{fig:2D_scheme}a) is excited with a sequence of three to four optical pulses (Fig.\,\ref{fig:2D_scheme}b) and the third-order nonlinear response of the system is probed as a function of the pulse delays $\tau,\, T$ and $t$. The time intervals $\tau$ and $t$ (termed coherence times), track the evolution of induced electronic coherences. A Fourier transform with respect to these time variables yields the 2D frequency-correlation maps (Fig.\,\ref{fig:2D_scheme}c) as parametric function of the third time variable $T$ (termed evolution time). Consequently, pump and probe steps are both frequency-resolved with $\omega_\tau$ representing the pump/excitation frequency and $\omega_t$ the probe/detection frequency axis, respectively. 

The detected signals are categorized in stimulated emission (SE), ground state bleach (GSB) and exited state absorption (ESA) each occurring as rephasing (RP) (photo echo) and non-rephasing (NRP) signals. GSB probes the time evolution on the system's ground state manifold whereas SE and ESA probe the dynamics of the excited state. Thereby ESA involves the excitation to a higher-lying state (Fig.\,\ref{fig:2D_scheme}). In most cases, GSB and SE pathways appear as positive and ESA as negative signals in the 2D maps which simplifies their identification and separation.

Furthermore, peaks on the diagonal reflect the linear absorption/emission spectrum of the sample, however, with the additional information of 2D lineshapes readily dissecting homogeneous (along antidiagonal) from inhomogeneous (along diagonal) broadening\,\cite{lazonder_easy_2006}. This provides decisive information about static and dynamic inhomogeneities in the probed ensemble\,\cite{maiuri_electronic_2018}. Off-diagonal features directly disclose couplings among excited states of the system from which different types of interaction and relaxation dynamics can be inferred, e.g.\ coherent excitonic interactions or spontaneous decay pathways\,\cite{lewis_probing_2012}. Excited state absorption (ESA) to higher lying states may be also induced. These contributions typically appear with inverted (negative) amplitude which simplifies their identification and separation from other contributions. All this information is in most cases hard to retrieve from one-dimensional spectra (Fig.\,\ref{fig:2D_scheme}d), indicating the great advantage of 2D spectroscopy. 

Furthermore, due to the Fourier transform-concept of CMDS, the time-frequency resolution automatically adapts to the system's time scales and spectral line widths\,\cite{jonas_two-dimensional_2003}. As such, broad bandwidth, ultrashort femtosecond pulses can be used to yield simultaneously high temporal and frequency resolution down to the Fourier limit, while probing transitions and correlations in a large spectral range\,\cite{ma_broadband_2016, kearns_broadband_2017}. 

\subsection{Technical challenges}
2DES faces two major technical challenges. First, tracking the femto- to sub-femtosecond beats of electronic coherences (during intervals $\tau,\, t$) requires interferometric measurements with high phase/timing stability among the optical pulses (typically $\leq \lambda / 50$\,\cite{hamm_concepts_2011}). This demand is slightly relaxed in 2DIR spectroscopy, since vibrational coherences evolve on roughly an order of magnitude lower frequencies. Second, the third-order 2D signals, subject to three to four light-matter interactions, are often weak and are covered by dominating background contributions, e.g. the linear system response or scattered light. This calls for highly sensitive detection with large dynamic range. 

In the past 25 years, both issues have been experimentally solved. A number of active and passive phase stabilization concepts have been developed to meet the demands of interferometric stability\,\cite{tian_femtosecond_2003, brixner_phase-stabilized_2004, grumstrup_facile_2007, vaughan_coherently_2007, tekavec_fluorescence-detected_2007, selig_inherently_2008, bristow_versatile_2009, turner_invited_2011, rehault_two-dimensional_2014, draeger_rapid-scan_2017}. These are combined with phase matching\,\cite{hybl_two-dimensional_1998} or phase cycling\,\cite{tian_femtosecond_2003, tan_theory_2008} schemes or combinations of both\,\cite{fuller_pulse_2014, dostal_direct_2018} to select the desired nonlinear signal contributions and provide highly sensitive background-free detection. An overview of the different experimental techniques has been recently published\,\cite{fuller_experimental_2015}. 

\begin{figure}
\centering
  \includegraphics[width=0.7\linewidth]{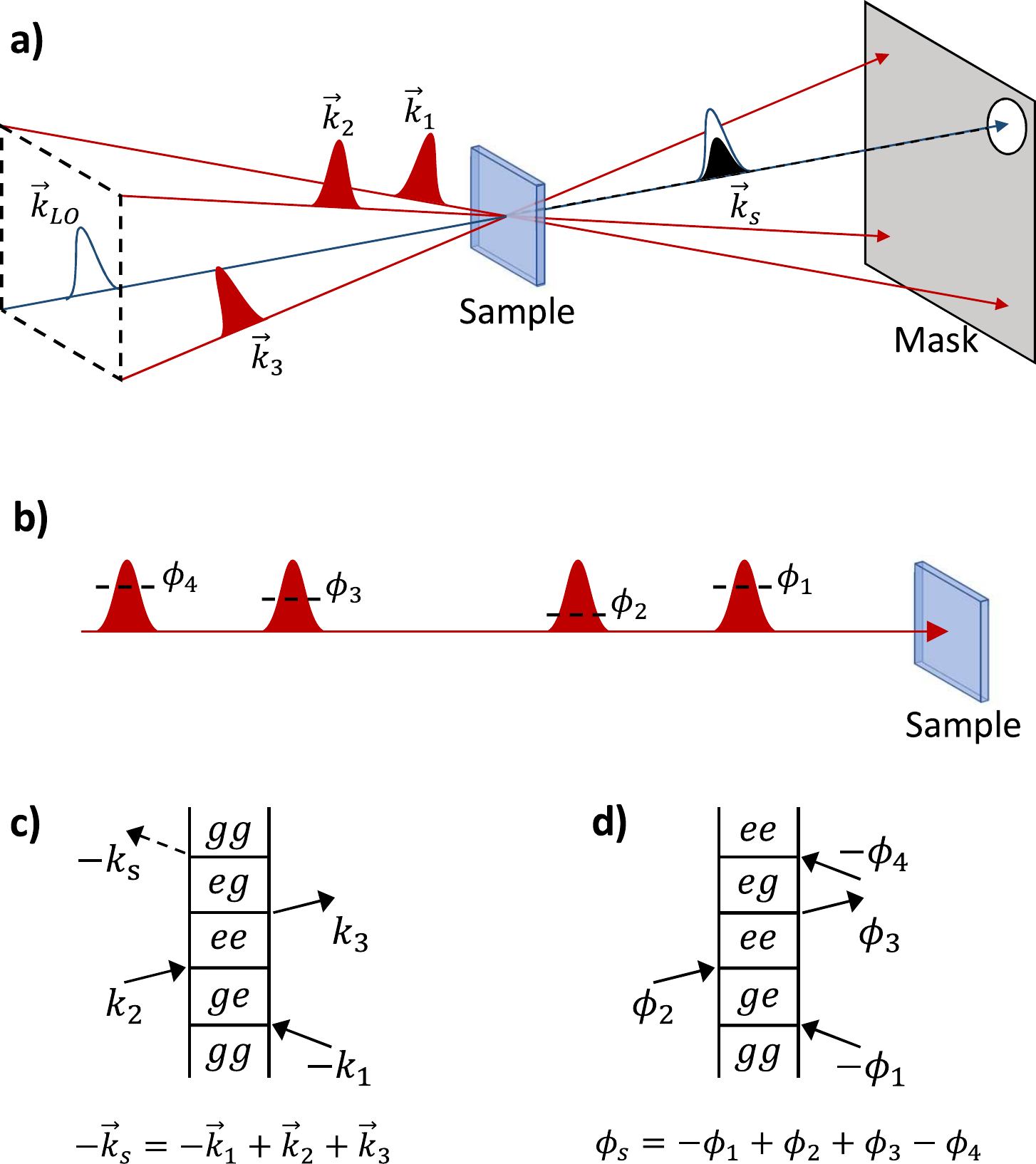}
  \caption{Phase matching and phase cycling in 2D spectroscopy. (a) Phase matching of three incident pulses with different $\vec{k}_i$-vectors induce a nonlinear polarization in the sample which radiates off in $\vec{k}_s$-direction where it is heterodyned with the local oscillator (LO) and isolated with a mask. (b) Phase cycling with four collinear pulses exciting a sample. The phase $\phi_i$ of each pulse is modified throughout the experiment. (c), (d) Example Feynman diagrams showing signal contributions selected with phase matching/cycling, respectively. 
  }
\label{fig:MatchingCycling}
\end{figure}
Phase matching (Fig.\,\ref{fig:MatchingCycling}a) relies on coherent four-wave-mixing (FWM), where the sample is excited with three laser pulses in the so-called boxcar geometry\,\cite{jonas_two-dimensional_2003}. The third light-matter interaction stimulates the coherent emission of the signal wave in phase matching direction, where it is background-free detected and frequency resolved with an optical spectrometer. Thereby, amplitude and phase of the signal are determined by heterodyned detection with a fourth optical field (termed local oscillator)\,\cite{hamm_concepts_2011}. 

In phase cycling (Fig.\,\ref{fig:MatchingCycling}b), collinear pulse trains are used to induce four light-matter interactions, leaving the sample in a population state after the fourth pulse. The final population is detected with incoherent observables yielding the nonlinear response of the system. At the same time, specific phase patterns are imprinted on the pulse trains by manipulating the carrier envelope phase (CEP) of each pulse, which results in a distinct phase signature of the detected signal\,\cite{tan_theory_2008}. By applying a unique set of phase combinations (typically 16 or 27), the desired nonlinear signal is identified and isolated in the post processing while other contributions destructively cancel. Here, the signal's amplitude and phase are deduced by adequate combination of extracted signal contributions.

The development of these concepts has solved some important technical issues of 2D spectroscopy experiments, having in recent years paved the way for widespread implementation in the condensed phase. Yet, other experimental issues exist that are less discussed in literature. These include timing uncertainties due to chirped optical pulses\,\cite{tekavec_effects_2010}, pulse overlap effects due to finite pulse durations\,\cite{perlik_finite_2017}, incomplete spectral overlap with the sample\,\cite{de_a._camargo_resolving_2017}, laser intensities beyond the weak perturbation regime\,\cite{chen_nonperturbative_2017}, pulse propagation effects in the studied medium itself\,\cite{li_pulse_2013, spencer_pulse_2015}, photo bleaching of samples and scattering light contributions\,\cite{augulis_two-dimensional_2011}. These points make 2DES still a sophisticated experimental task requiring specialized expertise in ultrafast nonlinear optics and related fields. 

\section{Experimental implementation of gas-phase 2DES}

\subsection{Action-based 2D spectroscopy}
The idea of gas-phase 2DES is to study isolated model systems, which implies very low ensemble concentrations (typically particle densities $\leq 10^{11}$\,cm$^{-3}$\,\cite{hertel_ultrafast_2006}). This requires orders of magnitude higher detection sensitivity than in condensed phase experiments (Fig.\,\ref{fig:overview}) and thus poses a severe technical challenge. The phase matching approach is ruled out by this criteria, as it relies on the coherent emission from a macroscopic polarization induced in the sample. Therefore, the method cannot be scaled down to low target densities and, to the best of our knowledge, phase-matching 2DES experiments have not been  demonstrated for particle densities $\leq 10^{12}$\,cm$^{-3}$\,\cite{dai_two-dimensional_2012}. 

In contrast, the phase cycling concept relies on the detection of a specific phase signature encoded in a nonlinear population state excited in individual particles. The respective nonlinear signal is deduced by mapping the population state with action-based detection of incoherent observables, e.g. spontaneous emission, depletion or photoionization. This approach does not rely on a macroscopic ensemble effect and, in principle, may be scaled down to the single-molecule level\,\cite{brinks_visualizing_2010}. As such, phase cycling has facilitated the development of action-based 2D spectroscopy, which opened a plethora of possibilities to incorporate new detection types. The combination with fluorescence\,\cite{tekavec_fluorescence-detected_2007, de_two-dimensional_2014, draeger_rapid-scan_2017}, photocurrent\,\cite{nardin_multidimensional_2013, karki_coherent_2014, vella_ultrafast_2016}, ion-mass\,\cite{roeding_coherent_2018, bruder_coherent_2018} as well as with optical microscopy\,\cite{goetz_coherent_2018, tiwari_spatially-resolved_2018} and even with high resolution photoemission electron microscopy\,\cite{aeschlimann_coherent_2011, aeschlimann_perfect_2015} has been demonstrated. 

\subsection{Pulse shaping versus continuous phase modulation}
Experimentally, phase cycling is implemented by pulse shaping based on spatial light modulators (SLMs)\cite{weiner_femtosecond_2000, vaughan_coherently_2007, turner_invited_2011} or acousto-optical modulators (AOMs)\,\cite{tian_femtosecond_2003, draeger_rapid-scan_2017, seiler_coherent_2017}. 
Alternatively, a phase modulation (PM) technique based on continuous phase modulation with acousto-optical frequency shifters (AOFSs) is used\,\cite{tekavec_fluorescence-detected_2007, nardin_multidimensional_2013, karki_coherent_2014, bruder_phase-modulated_2015, vella_ultrafast_2016, bruder_coherent_2018, Yu_2019}. 
The latter may be regarded as shot-to-shot quasi-continuous phase cycling\,\cite{nardin_multidimensional_2013}. 

Both approaches have their strengths and weaknesses depending on the application. Pulse shaping can drastically simplify the optical setups for 2DES experiments\,\cite{draeger_rapid-scan_2017, seiler_coherent_2017} and provide highest experimental flexibility. Amplitude, phase and polarization shaping permit the generation of arbitrary pulse sequences to perform a vast array of nonlinear spectroscopy experiments with a single apparatus\,\cite{brixner_femtosecond_2001}. Another advantage of pulse shapers is their ability for inherent pulse compression to yield transform-limited pulses in the 10-fs-regime\,\cite{goetz_coherent_2018}. 

On the contrary, the PM approach requires larger assemblies of optics and is more restricted in the manipulation of pulse properties. Yet, flexible signal selection protocols have been also implemented with the PM technique\,\cite{bruder_efficient_2015, Yu_2019} and pulse durations $<20$\,fs have been reported\,\cite{vella_ultrafast_2016}. The continuously operated AOFSs in the PM technique have the advantage of imprinting particularly clean, high purity phase manipulation with very low distortion (reported artifacts $\leq 50$\,dB\,\cite{lai_nonlinear_2014, bruder_phase-modulated_2017}), whereas pulse shapers require careful calibration and may produce artifacts due to space-time couplings\,\cite{frei_space-time_2009}, thermal phase instabilities\,\cite{bruhl_minimization_2017} or if operated at high update rates. 

In view of gas-phase experiments, the signal-to-noise performance and detection efficiency are particularly important factors. In 2DES, it is recommended to use moderate laser intensities to avoid the contribution of higher-order (larger than third order) signals to the data. Therefore, large statistics is best reached with low laser intensities and high laser repetition rates. Here, the PM technique has the clear advantage of providing shot-to-shot phase manipulation up to laser repetition rates in the MHz-regime\,\cite{tekavec_fluorescence-detected_2007} which is combined with highly sensitive lock-in detection\,\cite{tekavec_fluorescence-detected_2007, bruder_efficient_2015, bruder_delocalized_2019}. Most pulse shapers are restricted to update rates of $\leq 1$\,kHz\,\cite{draeger_rapid-scan_2017}. However, with the recent development of pulse shapers extending update rates to 100\,kHz \,\cite{kearns_broadband_2017}, the gap to the PM technique may be closed. 

Furthermore, gas-phase experiments may require much larger scanning ranges of pulse delays than typical in the condensed phase, where perturbations by the environment induce rapid dephasing of electronic coherences often within $\leq 100$\,fs. In the gas phase, broadening effects are considerably smaller and electronic coherences can be detected over hundreds of picoseconds\,\cite{bruder_phase-modulated_2015, lomsadze_frequency_2017}, enabling high resolution experiments. To exploit this feature in 2DES, coherence times have to be scanned over large time intervals, accordingly, which is possible in the PM approach\,\cite{bruder_phase-modulated_2015} but not with most pulse shapers where pulse delays are constrained to $\leq 1$\,ps\,\cite{draeger_rapid-scan_2017}. 

In our group, we favored the PM concept for the realization of gas-phase 2DES and we will in the following describe its experimental scheme in more detail. We note, that Brixner et al. realized gas-phase 2DES based on pulse shaping technology and we refer to their work for more information\,\cite{roeding_coherent_2018}. 

\subsection{Phase modulation 2DES combined with photoionization}
Phase modulation 2DES combined with fluorescence detection is described in detail in the original publication from the Marcus group\,\cite{tekavec_fluorescence-detected_2007}. Here, we provide only a brief description of the technique with the focus on the photoionization gas-phase experiments performed in our laboratory. 

\begin{figure*}
	\centering
	\includegraphics[width=0.7\linewidth]{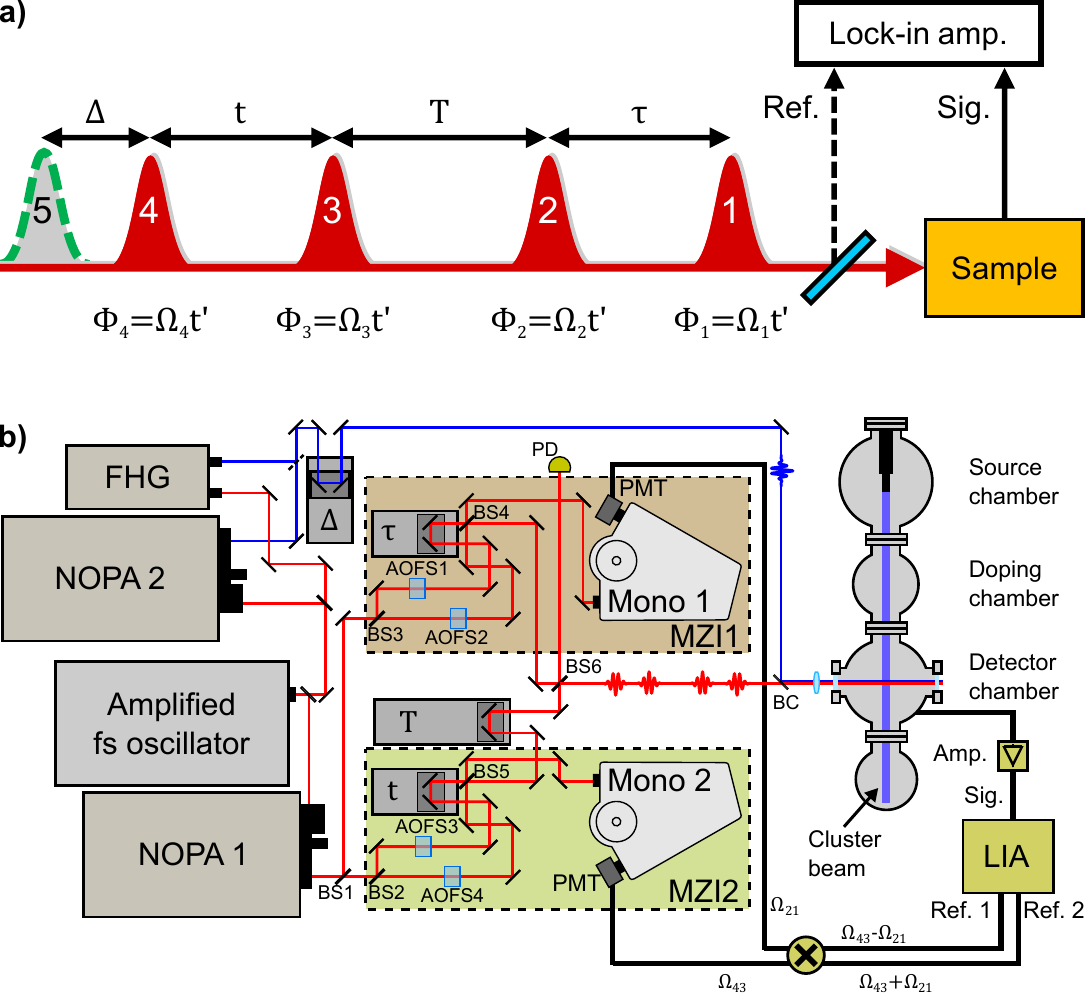}
	\caption{Detection scheme and optical setup for phase modulation 2DES in the gas-phase combined with photoionization. (a) A sample is excited by a collinear pulse sequence consisting of four phase-modulated pulses (pulse 1-4) and an optional fifth pulse (indicated by dashed envelope). The phase modulation of pulse 1-4 appears as characteristic beat notes in the detected nonlinear signals which are demodulated with a lock-in amplifier. A reference signal is constructed from the optical pulses for the lock-in demodulation. (b) Experimental setup. Three-fold Mach-Zehnder interferometer (MZI) equipped with four Acousto-optical frequency shifters (AOFSs) produces a collinear 4-pulse sequence (red). A fifth pulse (blue, not modulated) is collinearily overlapped with pulse 1-4 before focusing into the vacuum apparatus. Replicas of pulse pairs 1,2 and 3,4 are picked-up after beam splitter (BS) 4 and 5. Their low-frequency beats at $\Omega_2-\Omega_1=\Omega_{21}$ and $\Omega_{43}$ are detected in two monochromators (Mono 1,2) to construct the $(\Omega_{43}\pm \Omega_{21})$ sideband references (Ref. 1, 2) for the lock-in detection. $\tau,\,T\, ,t$, $\Delta$: pulse delays, NOPA: noncollinear optical parametric amplifier, FHG: fourth harmonic generation, BC: beam combiner, PMT: photo multiplier tube. Adapted from Ref.\,\cite{bruder_coherent_2018}, licensed under the \href{https://creativecommons.org/licenses/by/4.0/}{Creative Commons Attribution 4.0 International License.} 
	}
	\label{fig:PMsetup}
\end{figure*}

The experimental scheme and a sketch of the setup is shown in Fig.\,\ref{fig:PMsetup}. A collinear pulse train of four phase-modulated laser pulses prepares a nonlinear population state in the sample, which is probed upon photoionization. The ionization is either performed with a separate fifth pulse or by absorbing additional photons from pulse 4. 
Pulse 1-4 are generated in a nested three-fold optical interferometer fed by the output of a noncollinear optical parametric amplifier (NOPA) (640-900\,nm tuning range). Pulse 5 is produced from a second NOPA to enable independent wavelength tuning (540-900 \,nm) or from fourth harmonic generation (FHG) of the amplified oscillator pulses to yield deep ultraviolet (UV) pulses (260nm). The relative pulse delays are controlled by motorized translation stages. 

The multipulse excitation sequence generally induces a large number of signals. The desired third-order rephasing (RP) and non-rephasing (NRP) signal contributions are selected from the total signal by phase modulation of the excitation pulses combined with lock-in detection. To this end, pulse\,1-4 are passed through individual AOFSs (AOFS 1-4, Fig.\,\ref{fig:PMsetup}b) which are phase-locked driven at distinct radio frequencies $\Omega_i$. AOFS 1-4 shift the frequency of transmitted pulses by the value $\Omega_1 = 109.995$\,MHz, $\Omega_2 = 110.000$\,MHz, $\Omega_3 = 110.001$\,MHz and $\Omega_4 = 110.009$\,MHz, respectively. This is equivalent to a shot-to-shot modulation of the CEP $\phi_i$ of each pulse\,\cite{nardin_multidimensional_2013} in increments of $\Delta \phi_i = \Omega_i / \nu_\mathrm{rep}$ between consecutive laser shots ($\nu_\mathrm{rep}=200\,$kHz denotes the laser repetition rate). 

The nonlinear mixing of the modulated electric fields in the sample leads to characteristic beat notes in the photoionization yield. According to the phase cycling conditions for RP and NRP signals ($S_\mathrm{RP}$ and $S_\mathrm{NRP}$), the modulation frequencies are: 
\begin{eqnarray}
	S_\mathrm{RP} :  \phi_\mathrm{RP} (t) &=& -\phi_1 + \phi_2 + \phi_3 - \phi_4 = 3\,\mathrm{kHz}\\
	S_\mathrm{NRP} :  \phi_\mathrm{NRP} (t) &=& -\phi_1 + \phi_2 - \phi_3 + \phi_4 = 13\,\mathrm{kHz} \, .
\end{eqnarray}

The signals are extracted from the photoelectron/-ion count rates with lock-in detection. For the lock-in amplification, an external reference signal is used constructed from the optical interference of pulse 1-4. For this purpose, pulse pairs 1,2 and 3,4 are split off at BS\,4 and 5, respectively and are subsequently stretched in time with a monochromator (Fig.\,\ref{fig:PMsetup}b). The pulse stretching ensures a non-vanishing interference signal over sufficiently long scanning ranges of $\tau$ and $t$. The acquired beat signals of both pulse pairs ($\Omega_{21} = 5$\,kHz and $\Omega_{43} = 8$\,kHz) are electronically mixed to yield sum- and difference frequency-sidebands at 3 and 13\,kHz, respectively. 

In the post processing, the sum of demodulated RP and NRP signals is Fourier transformed with respect to the time delays $\tau$ and $t$ to yield the complex-valued 2D frequency-correlation spectrum $\tilde{S}(\omega_\tau, T, \omega_t)$ as a parametric function of $T$. Its real part represents the 2D absorption spectrum which is analyzed in the experiments. 

\begin{figure}
	\centering
	\includegraphics[width=0.8\linewidth]{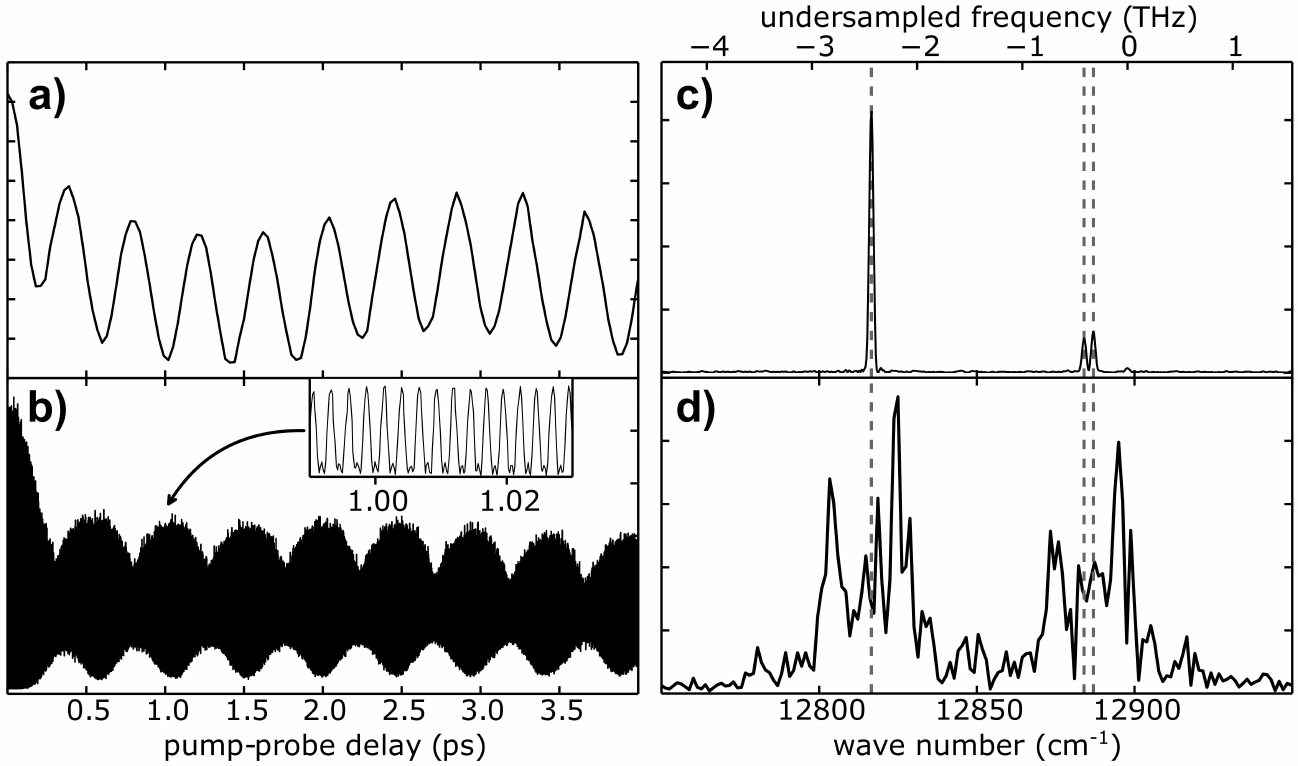}
	\caption{Performance advantage of the PM technique demonstrated in a quantum beat experiment. (a), (b): Time domain signal of electronic coherences excited in gaseous Rb atoms, with (a) and without (b) using the PM technique. In (a), rotating frame sampling leads to a downshift of the quantum beat frequencies. (c), (d): Respective Fourier transform spectra showing a clear SN advantage for the PM case. Adapted from Ref.\,\cite{bruder_phase-modulated_2015} - Published by the PCCP Owner Societies. 
	}
	\label{fig:PMvsClassic}
\end{figure}

The here employed lock-in detection scheme has several advantages. RP and NRP signals are retrieved simultaneously in a single 2D scan of positive coherence times $\tau$ and $t$. Amplitude and phase of the signal are recovered through phase-synchronous lock-in detection. Heterodyning with the external reference leads to rotating frame sampling which reduces the required delay sampling points by several orders of magnitude. Phase/timing jitter introduced in the optical interferometers appears as correlated noise in the signal and reference and thus cancels out in the lock-in demodulation, resulting in a highly efficient passive phase stabilization of the setup. As such, a phase stabilization better than $\lambda /200$ has been achieved in a deep-UV interferometer ($\lambda = 266$\,nm)\,\cite{wituschek_stable_2019}. Eventually,  the lock-in amplification considerably improves the general sensitivity of the setup. The signal-to-noise (SN) advantage of the PM technique is clearly demonstrated in an electronic quantum interference measurement combined with photoionization which served as a precurser experiment to our gas-phase 2DES experiments (Fig.\,\ref{fig:PMvsClassic})\,\cite{bruder_phase-modulated_2015}.

\subsection{Pathways in photoionization-2DES}

\begin{figure*}
\centering
  \includegraphics[width=0.9\linewidth]{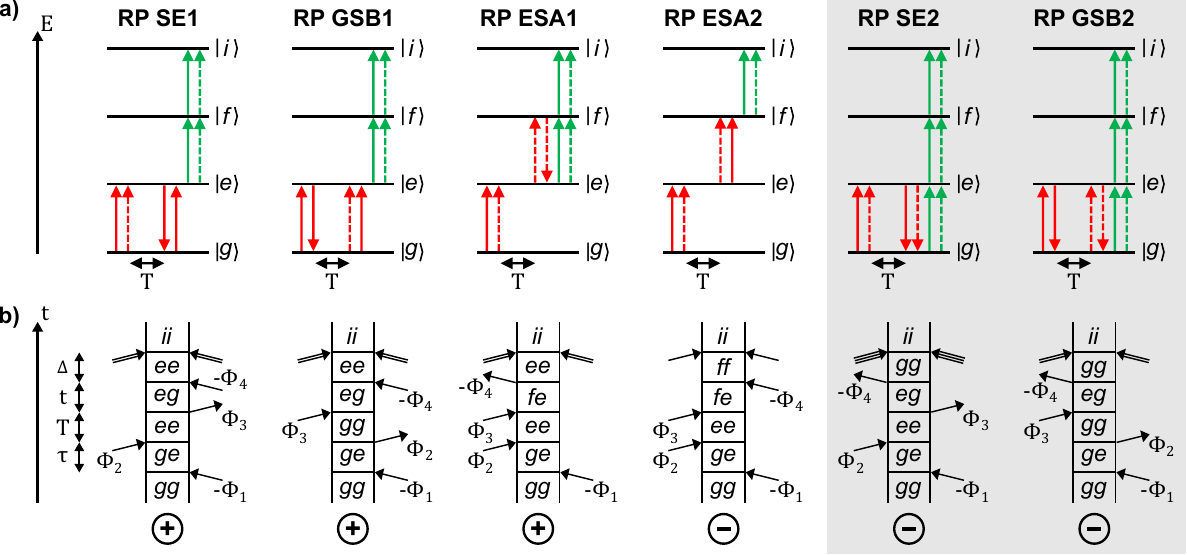}
  \caption{RP excitation pathways in photoionization-2DES. 
  (a) Simplified level scheme with $\ket{g}$ ground, $\ket{e}$, $\ket{f}$ excited states and $\ket{i}$ ionic state along with SE, GSB and ESA excitation pathways. Interaction by pulse 1-4 (red), pulse 5 (green). Solid arrows indicate interaction on ket-side, dashed on bra-side of the system's density matrix operator. (b) Corresponding double-sided Feynman diagrams. Common notation is used: Time evolves from bottom to top. Each entry denotes an element of the density matrix $\ket{i}\!\bra{j}$. Arrows indicate the light-matter interaction leading to de-/excitation of the system. Double-arrows indicate two simultaneous interactions.  $\phi_i$ indicates the phase imprinted onto the signal by each interaction. Plus/minus signs below each diagram indicate the sign with which the processes add to the 2D response function. Adapted from Ref.\,\cite{bruder_coherent_2018}, licensed under the \href{https://creativecommons.org/licenses/by/4.0/}{Creative Commons Attribution 4.0 International License.} 
}
\label{fig:pathways}
\end{figure*}

There is a distinct difference between 2D spectroscopy experiments using phase matching and phase cycling. In case of phase matching, for each signal type (SE, GSB, ESA) exists one RP and one NRP pathway (and their complex conjugate). With phase-cycling, for each contribution exists an additional pathway whose signal is phase shifted by $\pi$. Example RP pathways, as detected in our photoionization experiments, are shown in Fig.\,\ref{fig:pathways}. Here, the relative amplitude with which the pathways contribute to the signal strongly depend on the ionization probability. While SE1- ESA1 pathways are probed by two-photon ionization, the ESA2 process requires only one photon to the continuum and therefore dominates the ESA signal in the photoionization 2D spectra. On the contrary, SE2 and GSB2 require three photons for ionization and are usually negligible. As such, the net SE and GSB signals contribute with positive amplitude, whereas the net ESA signal strictly appears with negative amplitude in the 2D absorption spectra. This is in analogy to phase matching based 2D spectroscopy where the ESA amplitudes are also of opposite sign to SE/GSB signals. 

The negative sign of ESA features in contrast to the other signal contributions, simplifies their identification, which is of advantage, in particular in congested spectra. Note that with other detection types the situation can differ. In fluorescence detection, the sign of the ESA peaks depends on the degree of quenching of fluorescence from the $\ket{f}$ state\,\cite{perdomo-ortiz_conformation_2012}.

\subsection{Phasing of 2D spectra}
Related to the phase shift among signal contributions is the general \textit{phasing} issue in 2D spectra\,\cite{gallagher_faeder_two-dimensional_1999, turner_comparison_2011}. The correct phase information can only be retrieved if the total phase of the complex-valued 2D response function $S(\tau,T,t)$ is correctly determined. Otherwise the absorptive and dispersive line shapes are not correctly separated in the 2D absorption spectrum leading to distorted or even inverted peak shapes which might be interpreted incorrectly. 

While phasing of the 2D signals is intricate in FWM-based 2D spectroscopy, it is much simpler in collinear 2D spectroscopy experiments. Pulse shaper setups are intrinsically phased through the calibration of the device. In the phase modulation approach, phasing is done by calibrating the phase offset between the signal and reference in the lock-in detection. To this end, at coherence times set to zero ($\tau = t = 0\,$fs), the phase of the demodulated RP/NRP signal is adjusted to zero through adjusting a global phase factor applied in the lock-in electronics or in the post processing\,\cite{tekavec_fluorescence-detected_2007}. With this procedure, phase shifts between the signal and the reference accumulated in the different electronic circuits of the setup are compensated. 

\begin{figure}
\centering
  \includegraphics[width=0.9\linewidth]{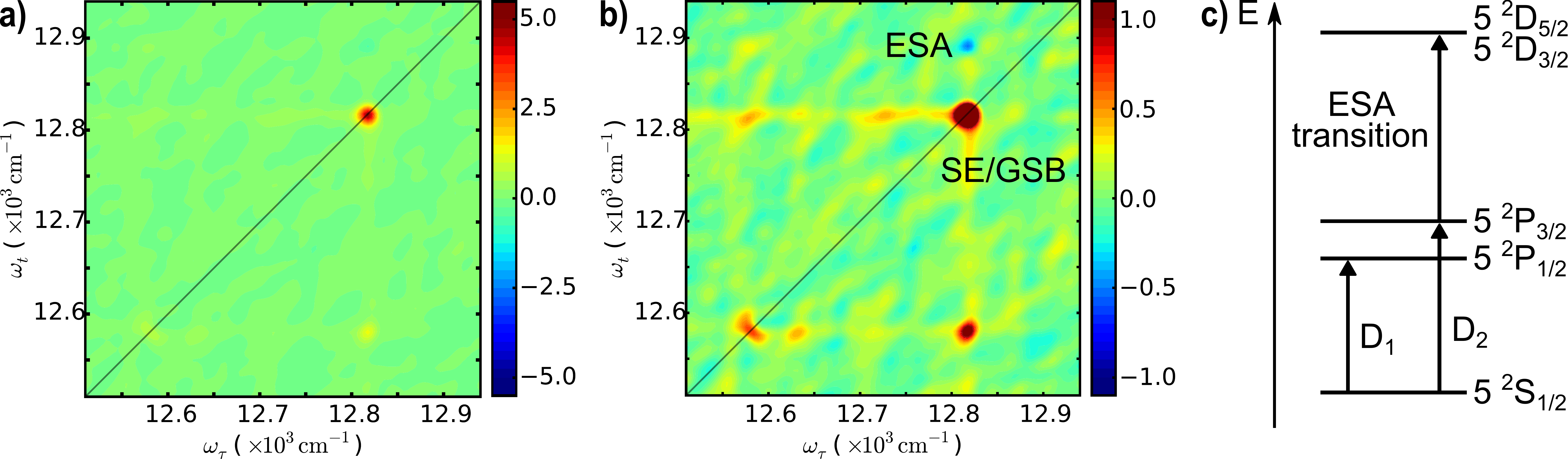}
	  \caption{Reference measurement to phase the setup. (a) Photoelectron-detected absorptive 2D-spectrum of gaseous Rb. Used to phase the setup according to the expected peak shape and sign: positive, absorptive GSB/SE features and negative, absorptive ESA features. (b) Different color-coding  to amplify weak contributions. Homogeneous/inhomogeneous broadenings are beyond the spectral resolution of the measurement, explaining the absence of peak elongations along the diagonal. (c) Relevant energy levels of atomic Rb for the reference measurement. Probed are the D line transitions (via GSB/SE) as well as the $5 {}^2P_{3/2} \rightarrow 5 {}^2D_{3/2, 5/2}$ transition (via ESA).}
  \label{fig:2Deffusiv}
\end{figure}

This procedure is required for the initial calibration of any PM-2DES setup, or whenever electronics are changed. Any reference sample may be used for the calibration. In our experiments, we phased the setup with photoionization signals of gaseous Rb atoms which provide a simple, well-defined 2D spectrum with isolated sharp peaks that allows for direct examination of any phase offset (Fig.\,\ref{fig:2Deffusiv}). 

\section{Preparation of gas-phase samples}

\subsection{Thermal vapors and molecular beams}

A great variety of experiments on atoms, molecules and molecular complexes are performed on gas phase samples, driven in particular by two main characteristics of such targets: (a) probing systems without the interaction between individual constituents or/and without the interaction with an environment; (b) establishing low temperature conditions and corresponding quantum state selectivity. 

With respect to (a), already gas cells containing a vapor pressure of the sample may evolve only weak perturbations in spectroscopic measurements. 
Albeit, the coherent excitation of molecular vapors (molecular densities $\sim 10^{18}$\,cm$^{-3}$) may lead to cascading effects which compromise the nonlinear response of the sample\,\cite{grimberg_ultrafast_2002}. 
Likewise, propagation effects may occur in gas cells\,\cite{li_pulse_2013, spencer_pulse_2015}. 
Experiments on particle beams prepared in high or ultra-high vacuum (UHV) environments circumvent these issues and in addition provide conditions  (pressures below  $\approx 10^{-5}$\,mbar) where the mean free path for extracting ions and electrons is suitable for an unperturbed detection. 
Furthermore, detection methods employing electron multipliers and corresponding high voltages cannot be operated at higher vacuum pressures.

With respect to (b), in the gas phase a variety of cooling and trapping methods are at hand to reach temperatures even down to nanokelvin temperatures (cf.\ Fig.\ \ref{fig:overview})\,\cite{krems2009cold}. A central technique is based on the cooling by means of a supersonic expansion in molecular beams \cite{scoles_atomic_1988}, reaching temperatures in the low Kelvin range. Ultracold temperatures (below mK) mostly involve laser cooling methods, as well as evaporative cooling in shallow traps \cite{krems2009cold}. Low temperatures are for many experiments instrumental for guaranteeing quantum state selectivity, preferable in all degrees of freedom, as well as providing well-defined structural properties. Finally, in comparison with ordinary gas targets, molecular beam as well as trapping methods are in many cases a prerequisite for providing an interaction volume having a distinct higher target density in comparison to the background gas inside the vacuum apparatus. Furthermore, Doppler broadening is minimized even in fast molecular beams when intersected perpendicularly by the laser beams. 

It is intriguing that independent of the very different experimental techniques providing gas-phase targets, like e.g.\ size-selected molecular or cluster ion beams, decelerated molecular beams\,\cite{van_de_meerakker_taming_2008}, helium droplet isolation, or ultracold atoms in magnetooptical traps (cf.\ Fig.\ \ref{fig:overview}), the target density typically is in the range of about $10^8$\,cm$^{-3}$. Of course, such densities are many orders of magnitudes below corresponding bulk target densities. However, the sensitivity and selectivity of signals detecting angular resolved and energy resolved single electrons or mass-selected ions even in sophisticated coincidence methods, in combination with generally fast regenerating targets offer unique options of experimental techniques not being available on bulk liquid or solid systems.

\begin{figure}
	\centering
	\includegraphics[width=0.7\linewidth]{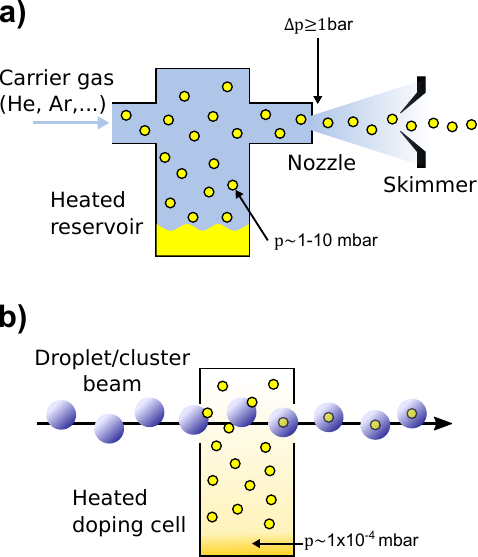}
	\caption{Gas-phase sample preparation. (a) Skimmed seeded supersonic beam generation. (b) Cluster isolation technique.}
\label{fig:Beams1}
\end{figure}

\begin{figure*}
	\centering
	\includegraphics[width=0.9\linewidth]{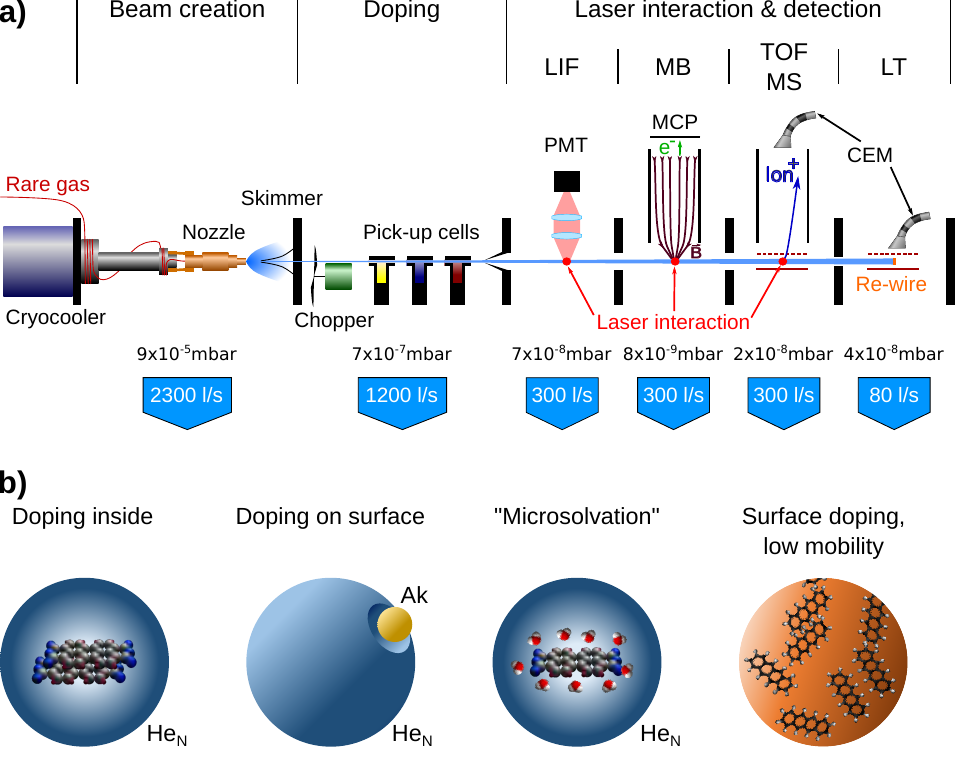}
	\caption{Helium nanodroplet isolation technique. (a) HENDI beam apparatus for 2DES measurements. Three different detection schemes can be used to gain complementary information: laser-induced fluorescence detector (LIF), magnetic bottle-type electron time-of-flight spectrometer (MB), ion time-of-flight mass spectrometer (TOF MS). A Langmuir-Taylor detector (LT) and a quadrupol-mass spectrometer (not shown) are used for beam diagnostics. PMT: photo multiplier tube. MCP: microchannel plate. CEM: channeltron electron multiplier. Typical pressure readings for the different chambers when operating the droplet beam, and pumping speed vacuum pumps are indicated below. (b) Doping of rare gas clusters. Most species immerse into the liquid He droplets and formation of larger atomic or molecular aggregates is possible. Co-doping of other atoms or molecules (microsolvation) allows for a precise tuning of the environmental parameters. Alkali (Ak) atoms and molecules are only weakly bound to the He droplet and reside on the droplet surface. Large clusters of Ne, Ar and Kr are solid and hence the dopants attach to the cluster surface exhibit a low mobility.}
\label{fig:Beams2}
\end{figure*}

The most commmon molecular beam technique is the generation of a skimmed seeded supersonic beam, (Fig.\, \ref{fig:Beams1}a). In an adiabatic expansion of high-pressure rare gases (He, Ar, Kr, Xe) into vacuum an  internally cold  beam traveling at supersonic speed is formed \cite{scoles_atomic_1988,Demtroeder_1988}, seeded with target molecules at much lower pressure, e.g. from a heated reservoir. In this way the molecules adapt in many collisions during the expansion process to the narrow speed distribution and the low temperature of the seed gas. In this way, both the directionality and density in the target volume is much higher, and the internal temperature is much lower in comparison with e.g.\ an effusive gas beam (molecules exiting a reservoir though a pin hole without collisions). 

In our first studies on 2DES in molecular beams we used the helium nanodroplet isolation (HENDI) technique, detailed out in the next section, because of its prospects and options for generating specific larger molecular structures at millikelvin temperatures.

\subsection{Cluster isolation technique}
Rare gas (Rg) clusters of variable sizes (Rg$_N, 1<N<10^{12}$) can be readily condensed in supersonic expansions at appropriate conditions. Depending on the rare gas, typically high stagnation pressures, $5-100$\,bar, and low temperatures, down to 4\,K  in case of He have to be applied\,\cite{Hagena1972,Hagena1987,Haberland1994}. Because of the low binding energies of rare gas atoms to the clusters and the high surface-to-volume ratio, the clusters very efficiently evaporatively cool to specific low temperatures. In helium, the terminal temperature is 380\,mK\,\cite{Hartmann:1995} which is well below the transition temperature to superfluidity. The liquid state and the superfluidity provides peculiar properties, in particular frictionless flow and efficient cooling which has been confirmed in many helium cluster studies, \cite{Grebenev:1998,Choi:2006} and explain why such clusters are appropriately called droplets. All rare gas clusters can be loaded with atoms and molecules by the pickup technique \cite{Lewerenz:1995,Toennies:1998}, where during inelastic collisions, e.g.\ in a cell containing a low vapor pressure of the dopant atoms or molecules, these are attached to the clusters. In comparison with seeded beams the needed partial pressure for doping a large cluster with unit probability is on the order of $10^{-5} - 10^{-4}$\,mbar, significantly extending the range of  molecules suitable for establishing such low densities without fragmentation. One can dope large clusters even with thousands of atoms or molecules \cite{Tiggesbaumker:2007}. A variety of doping techniques has been developed, including laser ablation\,\cite{claas_characterization_2003,mudrich_kilohertz_2007,katzy_doping_2016} and dopants from electrospray (ESI) sources\,\cite{bierau_catching_2010,filsinger_photoexcitation_2012,florez_ir_2015}. In this way, also charged particles have been doped. In combination with ion traps, cluster-isolated spectroscopy of large bio-molecules up to 12000\,Dalton has been performed\,\cite{bierau_catching_2010}. 

In helium, generally, dopants aggregate inside the liquid droplet and in this way one can specifically synthesizes even larger atomic or molecular structures (Fig.\ \ref{fig:Beams2}b) and/or model solvation effects by adding specific solvent molecules. On the other hand, the larger clusters of heavier rare gas atoms (Ne, Ar, Kr, Xe) all form solid clusters. For such solid clusters, it has been shown that larger molecules upon doping do not submerge and are immobile\,\cite{dvorak_spectroscopy_2012,dvorak_spectroscopy_2012-1}. In this way, multiple doping leads to a variable surface coverage of the doped molecules (cf.\ Fig.\ \ref{fig:Beams2}b)\,\cite{muller_cooperative_2015,izadnia_singlet_2017}.

\begin{figure}
	\centering
	\includegraphics[width=0.6\linewidth]{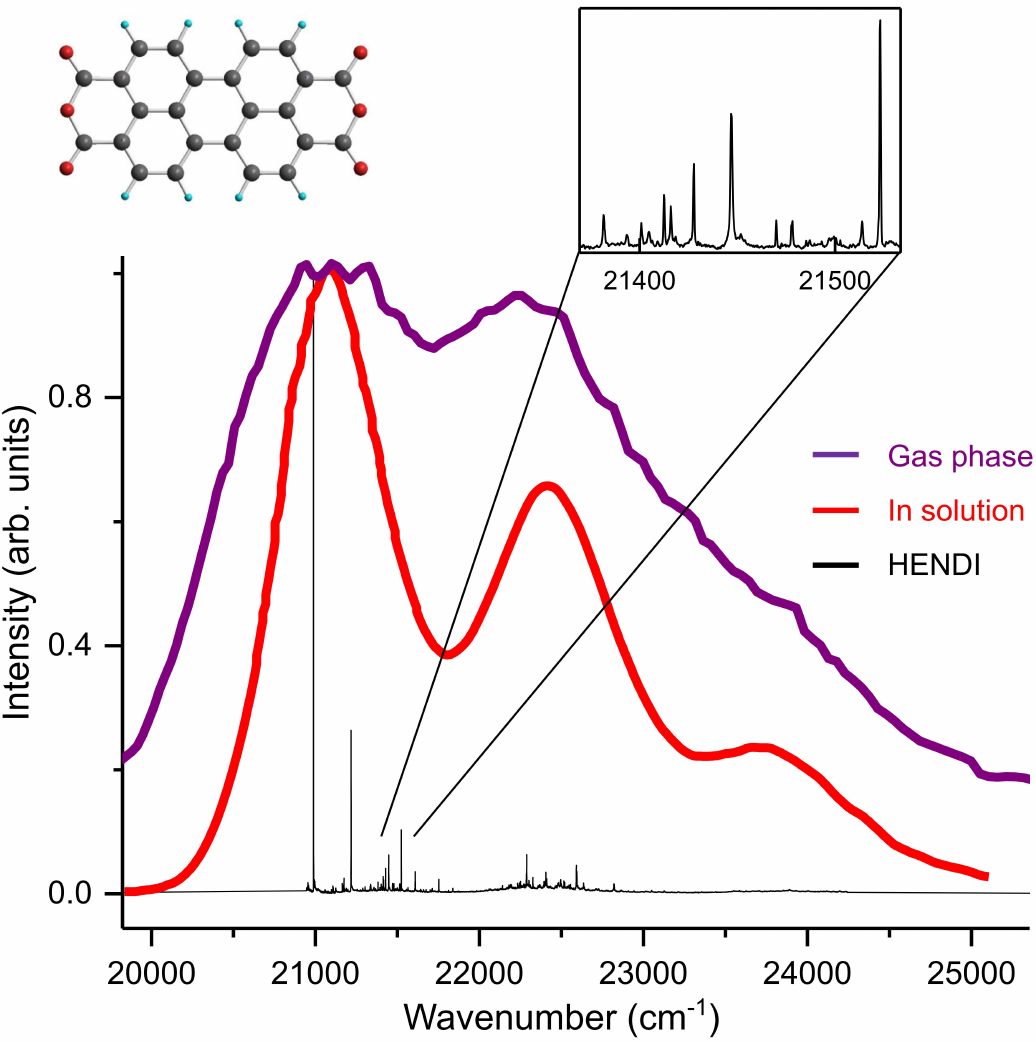}
	\caption{Comparison of linear absorption spectra of PTCDA (3, 4, 9, 10-perylenetetracarboxylicdianhydride) in different environments. Purple: gas-phase absorption in a heated vapor cell,\cite{Stienkemeier:2001}. Red: measurement in a room temperature solvent (dimethyl slufoxide)\cite{Bulovic:1996_2}. Black: helium droplet isolated monomer spectrum\,\cite{Wewer:2003}.}
\label{fig:PTCDA}
\end{figure}

All kinetic energy from the doping process as well as internal energy of the formed aggregates are dissipated via the evaporated cooling of the cluster. In this way, low temperature targets are formed, e.g., for helium droplets at millikelvin temperatures. Since the rare gas clusters are transparent at all wavelength down to the VUV, the dopants are selectively probed in laser experiments operating at IR, VIS or UV wavelengths\,\cite{Toennies:1998,Stienkemeier:2006}. 

Fig.\ \ref{fig:PTCDA} demonstrates the advantage of helium droplet isolation in the comparison of linear absorption spectra of PTCDA molecules at different conditions. The spectrum in a room temperature solvent shows the typical broad absorption bands of the S$_1 \leftarrow$ S$_0$ first singlet-to-singlet transition (red curve in Fig.\ \ref{fig:PTCDA}). Even the gas-phase absorption in a heated vapor cell  does not lead to better-resolved details (purple curve in Fig.\ \ref{fig:PTCDA}) because of the large number of thermally populated states. The helium droplet isolated spectrum, however, clearly resolves in detail the vibrational structure of the molecule. 

With the latter technique, the broadening of lines in vibronic spectra typically is about 1 cm$^{-1}$ \cite{Stienkemeier:2001}. The main source of broadening  often is the Pauli repulsion of the electron density with the surrounding helium. For atoms having low ionization potentials and  corresponding extended electron density distributions, large blue-shifts and  repulsive interactions may appear upon excitation of electronic states. The repulsive nature of helium with respect to electrons can even lead to the formation of so called ``bubbles'' \cite{Dalfovo:1994}, i.e.\ a helium void around e.g.\ atomic dopants. For the same reasons alkai atoms, dimers and trimers do not submerged in helium but are located at dimple-like structures on the surface of helium droplets (cf.\ Fig.\ \ref{fig:Beams2}b) having binding energies only on the order of 10\,cm$^{-1}$ \cite{Ancilotto1995,Stienkemeier:1996}.

\begin{figure}
	\centering
	\includegraphics[width=0.9\linewidth]{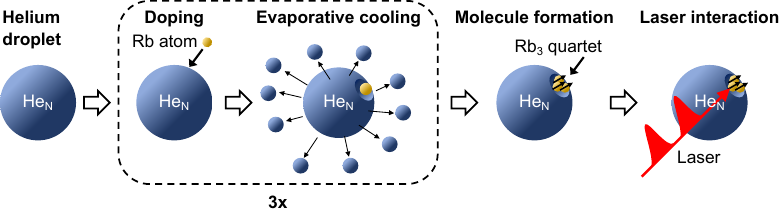}
	\caption{He droplet assisted formation mechanism of the investigated Rb molecules. By picking up several atoms, molecules are formed on the cluster surface. Evaporation of He atoms efficiently dissipates the released binding energy and cools the formed molecule to its vibrational ground state. Due to this mechanism, the formation of the Rb molecules in their lowest weakly-bound high-spin state is preferred. The higher binding energy of the low-spin electronic ground state molecules leads to desorption or droplet destruction, due to which these molecular configuaritons are normally not detected in the experiments. Further downstream, the prepared doped droplets are probed via 2DES. Graphic taken from Ref.\,\cite{bruder_coherent_2018}, licensed under the \href{https://creativecommons.org/licenses/by/4.0/}{Creative Commons Attribution 4.0 International License.} 
	}
\label{fig:Mol_formation}
\end{figure}

The just introduced peculiar binding properties of alkali-doped helium droplets preferably leads to the formation of high-spin states upon the formation of alkali molecules or clusters (Fig.\ \ref{fig:Mol_formation})\,\cite{stienkemeier_use_1995,Higgins:1998,Buenermann:2007}. Dissipation of binding energy upon the formation of molecules leads to high desorption rates of strongly bound entities during the doping process. In this way, in particular weakly bound molecules can be studied, which might be very difficult to form by other techniques. Alkali molecules in weakly-bound high-spin states have been probed in the first 2DES studies on helium droplets. 

During the last 20 years, helium droplet isolation has been applied to a large variety of spectroscopic techniques. The results have been reviewed in various publications and we refer to these for further information\,\cite{Callegari:2001,Stienkemeier:2001,toennies_superfluid_2004,Stienkemeier:2006,Choi:2006,Barranco:2006,Tiggesbaumker:2007,CallegariErnst:2011}.

\subsection{Helium nanodroplet beam apparatus}
A typical helium nanodroplet apparatus is depicted in Fig.\ \ref{fig:Beams2}a.  Helium droplets (He$_N$) with an average size of $N\approx 10000$ helium atoms per droplet form in a supersonic expansion at $P_0 = 50$\,bar stagnation pressure and about $T_0 = 15$\,K nozzle temperature. The molecular beam machine consists of a differentially pumped linear chain of HV/UHV vacuum chambers guiding the initially formed helium droplet beam via the doping unit to different detection chambers. Laser pulses can be introduced alternativly into a fluorescence detector, a magnetic bottle-type electron time-of-flight (TOF) spectrometer or a ion-TOF mass spectrometer, respectively. The mildly focussed laser and the droplet beam intersect perpendicularly. Since the droplet beam is travelling at about 400\,m/s and the repetition rate of the laser is 200\,kHz, each set of 2DES laser pulses acts on a fresh section of the target beam. Typical signal rates are one ion/electron per laser shot at target densities of about $10^{8}$ droplets per cm$^{-3}$. The magnetic bottle spectrometer for photoelectron spectroscopy has a resolution $\Delta E/E\approx 2$\,\% and includes retarder electrodes for shifting electron flight times. Further details of the machine used for 2DES can be found in other publications\,\cite{bruder_phase-modulated_2015}.

\section{Gas-phase 2DES of isolated, cold molecules}

\begin{figure}
\centering
  \includegraphics[width=0.95\linewidth]{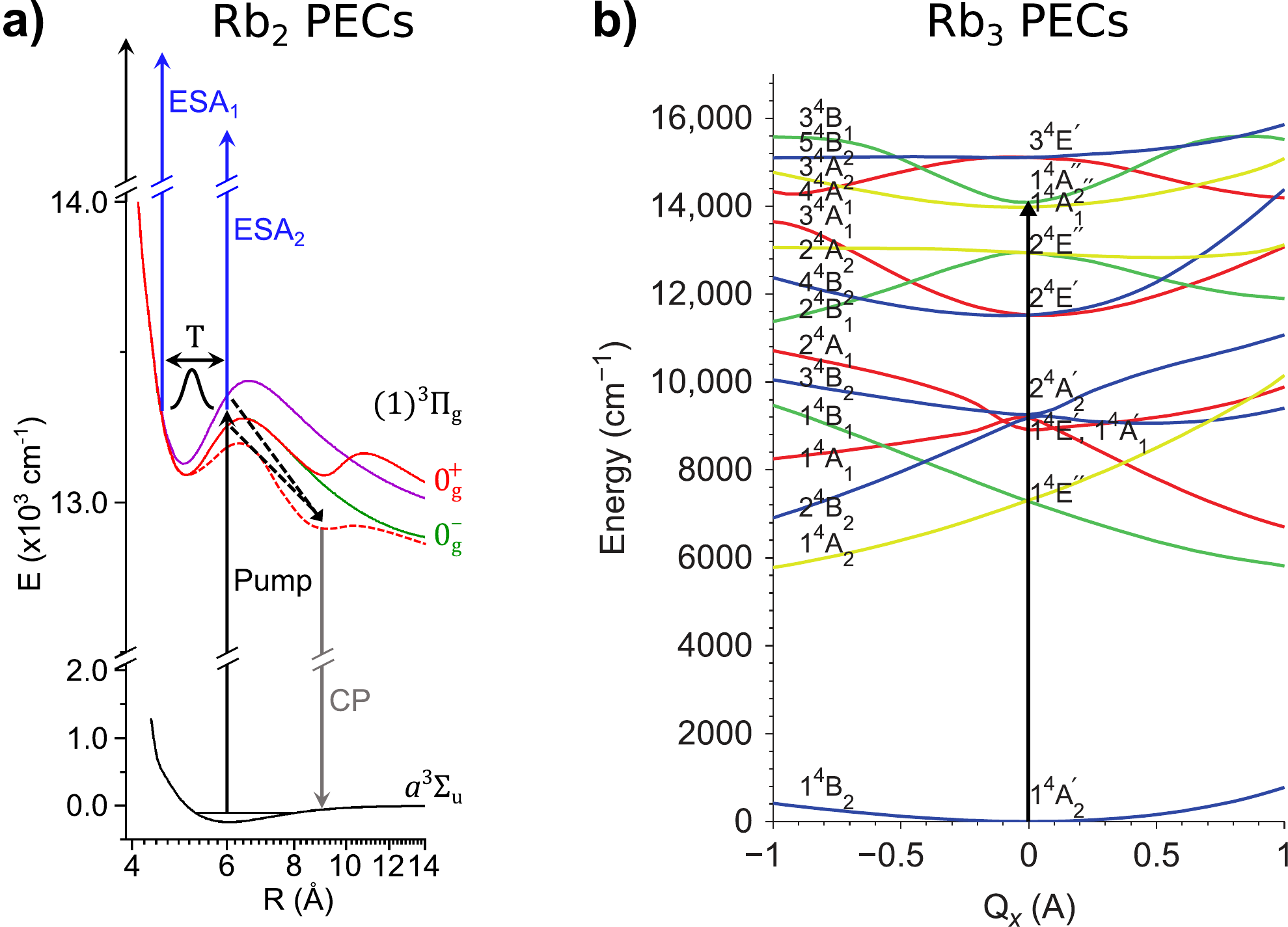}
  \caption{PECs of Rb$_2$ triplet (a) and Rb$_3$ quartet (b) manifolds. Arrows indicate the probed transitions. In (a), the perturbation of the $0_\mathrm{g}^+$ state by the helium environment is schematically shown as dashed curve. The Rb$_2$ PEC graphic is adapted from Ref.\,\cite{bruder_coherent_2018}, licensed under the \href{https://creativecommons.org/licenses/by/4.0/}{Creative Commons Attribution 4.0 International License}. The Rb$_3$ PEC graphic is adapted by permission from Springer Nature: Ref.\,\cite{hauser_advances_2009}, License Number: 4606941432839.}
\label{fig:PECs}
\end{figure}

\subsection{2DES of weakly-bound rubidium molecules}
\begin{figure*}
\centering
  \includegraphics[width=0.9\linewidth]{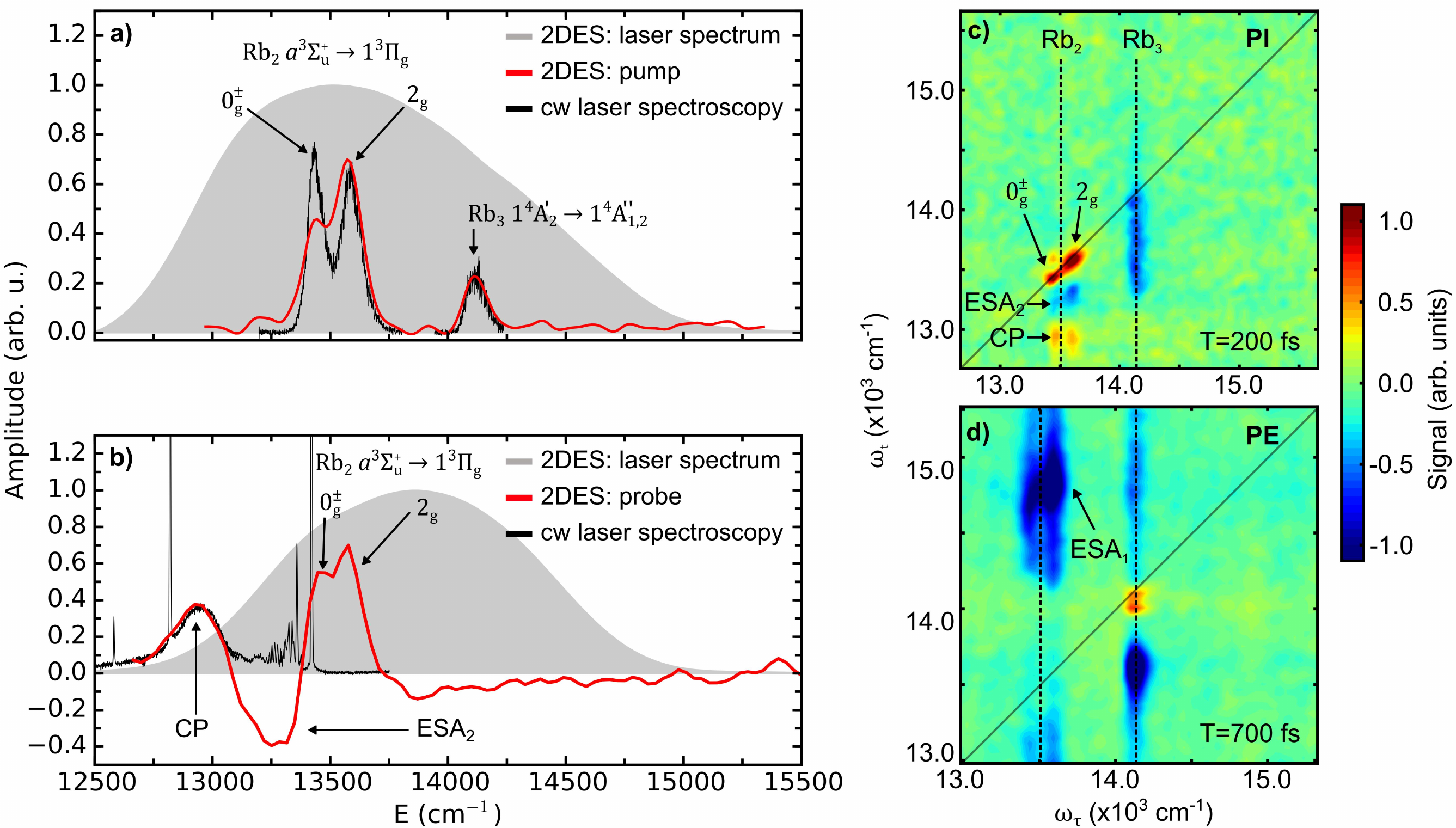}
  \caption{Comparison of Rb$_2$ and Rb$_3$ spectra obtained with 1D and 2D spectroscopy methods. (a,b) Absorption and emission spectra from high resolution laser spectroscopy (black) and 2D spectroscopy (red), obtained from a horizontal/vertical cuts through the 2D spectra in (d,c), respectively. The laser spectrum used in the 2DES experiments is shown in the background (gray). (c) Photoion-detected 2D spectrum at population time $T=200$\,fs, excitation pulse 1-4 center wavelength: 722\,nm, ionization pulse: 670\,nm. (d) Photoelectron-detected 2D spectrum at population time $T=700$\,fs, excitation pulse 1-4 center wavelength: 732\,nm, measured without additional ionization pulse. Color scale is saturated by a factor of 1.3. Adapted from Ref.\,\cite{bruder_coherent_2018}, licensed under the \href{https://creativecommons.org/licenses/by/4.0/}{Creative Commons Attribution 4.0 International License}.}
\label{fig:Rb2Rb32DES}
\end{figure*}

Recently, we have combined PM-2DES with HENDI and studied Rb$_2$ and Rb$_3$ molecules prepared in their weakly-bound high-spin states. These experiments constitute the first 2DES study of isolated, cold molecules prepared at sub-Kelvin temperatures. 

Fig.\,\ref{fig:PECs} shows the potential energy curves (PECs) of the molecules. Both molecules have been previously studied with HENDI using high resolution steady-state laser spectroscopy\,\cite{allard_investigation_2006, nagl_heteronuclear_2008, nagl_high-spin_2008} and femtosecond quantum beat spectroscopy\,\cite{mudrich_spectroscopy_2009, gruner_vibrational_2011, giese_homo-_2011}. The steady-state laser absorption and emission spectra are shown in Fig.\,\ref{fig:Rb2Rb32DES}a, b. The Rb$_2$ molecule shows a pronounced absorption at the $ 1^3\Pi_\mathrm{g}\leftarrow a^3\Sigma_\mathrm{u}^+ $ excitation with resolved spin-orbit (SO) couplings of the excited state. Note, that the $1_\mathrm{g}$ absorption does not appear in HENDI experiments. In the Rb$_3$ molecule, the $1^4 A_{1,2}^{''} \leftarrow 1^4 A_2^{'}$ absorption peak is observed. Emission spectra have been only reported for the Rb$_2$ molecule (Fig.\,\ref{fig:Rb2Rb32DES}b). 2D spectra taken of the same molecules attached to helium droplets are shown in Fig.\,\ref{fig:Rb2Rb32DES}c for photoion detection and in (d) for photoelectron detection. Both are taken under different ionization conditions with the purpose to selectively amplifying certain features (see discussion below). 

The 2D frequency-correlation maps exhibit high quality, in particular if considering the challenging experimental conditions. They show sharp, well-separated spectral features, which is not common in condensed phase studies, indicating the resolution advantage of the gas-phase approach. Remarkably, these spectra were taken for very small number densities of doped droplets being only $n \approx 10^7$cm$^{-3}$ which corresponds to roughly 300 absorbers inside the laser interaction volume. For these conditions, the integral optical density (OD) of the sample estimates to $\mathrm{OD}=-\log_{10}(I/I_0)\sim 10^{-11}$\cite{bruder_coherent_2018}. This is several orders of magnitude lower than in previous 2D spectroscopy studies, where the OD typically ranges between 0.1 and 1. Our experiments thus indicate a drastic improvement in sensitivity and open up new possibilities for an expansion to other fields, e.g. ultra cold atom clouds\,\cite{bloch_many-body_2008} and ion crystals\,\cite{gessner_nonlinear_2014} or towards single-molecule studies\,\cite{brinks_visualizing_2010}. 

In comparison with the previous 1D spectroscopy measurements of the molecules, the advantages and additional information gained by 2D spectroscopy become apparent. While the 2D spectra show the same absorption lines as in the 1D steady-state spectroscopy, correlated to the absorption bands, additional ESA pathways (negative peaks) and cross-peaks (red shifted positive peaks) are revealed. The ESA features expose the different ionization pathways as a function of the molecular excitation and show the position of respective Frank-Condon (FC) windows to higher-lying states\,\cite{bruder_coherent_2018}. This information was not available in previous photoionization studies of these molecules\,\cite{mudrich_spectroscopy_2009}, but, in principle, may be gained by narrow-band two-color pump-probe ionization experiments. The advantage of the 2DES approach is, that high spectral resolution is gained even when using broadband femtosecond pulses (see pulse spectra in Fig.\,\ref{fig:Rb2Rb32DES}a,b), that cover all transitions simultaneously and also permit femtosecond temporal resolution. 

Furthermore, in Fig.\,\ref{fig:Rb2Rb32DES}a, we show a direct comparison of line shapes obtained from steady-state spectroscopy and 2DES. To this end, the 2D spectra were integrated along certain horizontal/vertical spectral intervals\,\cite{bruder_coherent_2018}. We find a remarkably good match of absorption and emission profiles and equal spectral resolution for both methods, confirming that the Fourier transform concept of 2D spectroscopy indeed achieves optimum spectral resolution.  Note that the relative amplitude missmatch in the absorptive profiles of Rb$_2$ corresponds to different ionization probabilities of the respective states. 

For the Rb$_2$ emission spectrum, the situation is slightly different. The steady-state spectroscopy captures mainly the emitted fluorescence of free gas-phase molecules which tend to desorb from the droplet surface after their excitation\,\cite{vangerow_dynamics_2015, sieg_desorption_2016}. This explains the absence of the ESA resonance (negative peak at 13\,250\,cm$^{-1}$) and the well resolved vibronic features around 13\,300\,cm$^{-1}$. 

In contrast, the 2DES measurements provide spectral information with femtosecond time resolution (as a function of the evolution time $T$) and reproduce the spectrally broadened response of the Rb$_2$ molecules while being still attached to the droplet surface. As such, an almost identical pump and probe profile of the $a^3\Sigma_\mathrm{u}^+ \rightarrow 1^3\Pi_\mathrm{g}$ transition is observed. The absence of a Stokes shift is due to a very narrow FC window between the shallow ground state potential and the $1^3\Pi_\mathrm{g}$ state\,\cite{bruder_coherent_2018}. 

The femtosecond time resolution of the 2DES study, furthermore, has revealed the coherent WP dynamics of the Rb$_2$ molecule (not shown), which permitted a refined interpretation of the Stokes peak appearing at 12\,900\,cm$^{-1}$. While this feature was previously interpreted as the emission from vibrationally relaxed free gas-phase Rb$_2$ molecules\,\cite{allard_investigation_2006}, the 2DES experiments point to an ultrafast intramolecular relaxation into the outer potential well of the  $1^3\Pi_g$ state, catalyzed by the helium perturbation. This population transfer shows a remarkable efficiency, taking place within $< 100$\,fs\,\cite{bruder_coherent_2018}. 

\subsection{System-bath couplings in a superfluid environment}
\begin{figure*}
\centering
  \includegraphics[width=0.9\linewidth]{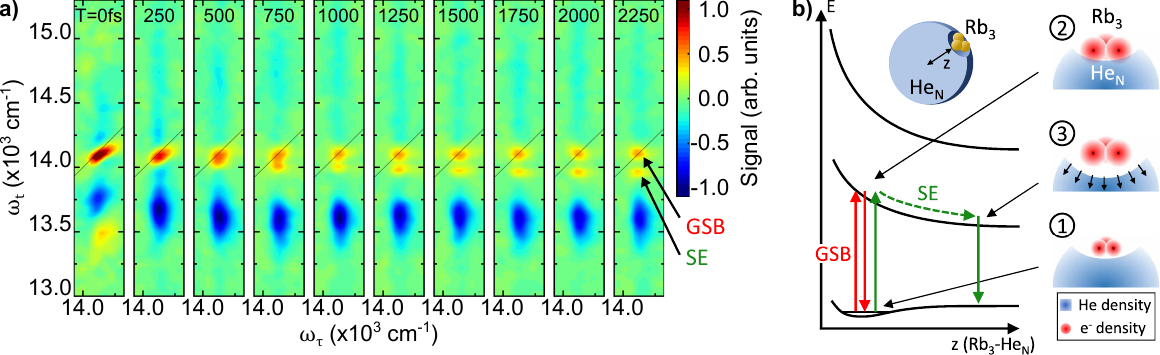}\vspace*{2mm}
  \caption{System-bath dynamics in the Rb$_3$He$_N$ system. (a) Time evolution of spectral features correlated to the Rb$_3$ $1^4 A_2^{'} \rightarrow 1^4 A_{1,2}^{''}$ absorption. (b) Schematic of the Rb$_3$-He$_N$ interaction potentials. Steps 1-3 sketch the repulsion of the helium density following the impulsive excitation of the Rb$_3$ molecule. The excited state relaxation process is traced by SE signals, whereas GSB signals probe the ground state where no system-bath dynamics occur. Adapted from Ref.\,\cite{bruder_coherent_2018}, licensed under the \href{https://creativecommons.org/licenses/by/4.0/}{Creative Commons Attribution 4.0 International License.}
  }
\label{fig:Rb3Stokes}
\end{figure*}
Using the helium nanodroplets as a matrix to isolate and cool down molecules, has the advantage of forming enclosed nanometer-sized model systems in an ultra high vacuum environment. This allows us to expand our studies beyond pure intramolecular dynamics towards intermolecular effects and the exploration of system-bath couplings induced by  well-controlled environmental parameters. 

For instance, different types of system-bath interactions can be modeled by co-doping of the droplets with rare gas atoms or solvent molecules (Fig.\,\ref{fig:Beams2}b)\,\cite{muller_cold_2009, dvorak_size_2014}. Alkali-metals show here an exceptional behavior and induce already a significant interaction with the pure helium droplets. This is explained by the strong Pauli-repulsion between the loosely-bound valence electrons of the alkali-metal atoms with the closed-shell $1s^2$ configuration of the helium atoms\,\cite{Buenermann:2007}. Alkali atoms and molecules thus serve as ideal probes to sense the interaction potentials and dynamical behavior of the superfluid droplets and allow us to explore the properties of the quantum fluid itself. 

An example of static guest-host interaction has been already discussed above for the perturbation of the Rb$_2$ $(1)^3\Pi_g,\,0_\mathrm{g}^+$ potential, opening up an ultrafast molecular relaxation channel which is not observed in the gas phase. The Rb$_3$ excitation reveals, on the contrary, an example of the ultrafast dynamic droplet response when impulsively pumping energy into the system through an impurity (Fig.\,\ref{fig:Rb3Stokes}). 

Upon excitation of the Rb$_3$ molecule, its electron density distribution expands, causing a repulsion of the surrounding helium atoms on a few-picosecond time scale, while the heavy Rb$_3$ molecule effectively remains in position and slowly desorbs from the droplet surface on a much longer time scale (estimated to be $>10$\,ps). 

The 2DES experiment allows us to directly follow the initial fast repulsion of the quantum liquid. 
Here, we observe a dynamic Stokes shift along the probe-frequency axis of the 2D spectra (Fig.\,\ref{fig:Rb3Stokes}a), which reflects the system's relaxation on the excited state of the Rb$_3$-He$_N$ interaction potential (Fig.\,\ref{fig:Rb3Stokes}b). Our time-resolved study reveals a rearrangement of the helium density towards the Rb$_3^*$He$_N$ equilibrium state within 2.5\,ps. Note, that the Rb$_3$ peak on the diagonal position reflects the dynamics on the system's ground state where the Rb$_3$-He$_N$ interaction is of static character. 

The here discussed example of system-bath dynamics represents a unique case where a single, isolated molecule interacts with a homogeneous environment. It allows us to directly probe the system-bath interaction potential without inhomogeneous broadening. This is in contrast to condensed phase studies, where a statistical ensemble of molecules is probed in an inhomogeneous environment. There, local bath fluctuations typically lead to a diffusion of the lineshape over time\,\cite{asbury_water_2004, moca_two-dimensional_2015} rather than resolving a dynamic Stokes shift. As such, our experimental approach provides an interesting alternative route to elucidate the influence of environmental parameters on molecular processes. 

\section{Photoionization as a selective and versatile probe}
One benefit of the photoionization is the vast array of highly developed electron/ion detectors enabling for instance energy- or mass-resolved spectroscopy. Depending on the detection method, one hereby is able to add extra dimensions to the 2D spectroscopy measurements which permits further disentanglement of excitation and reaction pathways.

\subsection{Coherent two-dimensional electronic mass spectrometry}
\begin{figure}
\centering
  \includegraphics[width=0.8\linewidth]{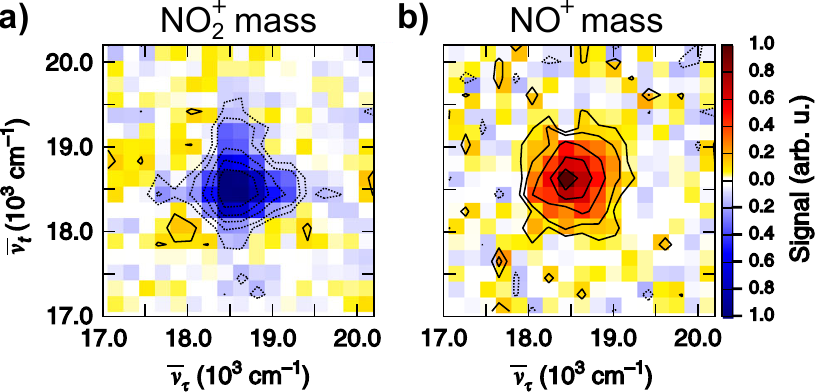}
  \caption{Ion-mass detected 2DES of gaseous NO$_2$ molecules. Absorptive 2D correlation spectra extracted from the NO$_2^+$ parent ion in a) and the ion fragment NO$^+$ in b). Adapted from Ref.\,\cite{roeding_coherent_2018}, licensed under the \href{https://creativecommons.org/licenses/by/4.0/}{Creative Commons Attribution 4.0 International License}.}
\label{fig:Brixner}
\end{figure}
A first demonstration of 2D spectroscopy combined with ion-mass detection has been recently reported by Brixner et al.\,\cite{roeding_coherent_2018}. Here, a warm effusive beam of gaseous NO$_2$ molecules was studied. Multiphoton excitation and ionization of the sample was induced by four collinear VIS pulses produced by an acousto-optical pulse shaper. Rapid shot-to-shot phase cycling was incorporated to improve signal to noise performance\,\cite{draeger_rapid-scan_2017} and to isolate the nonlinear response from NO$_2^+$ and  NO$^+$ ion yields (Fig.\,\ref{fig:Brixner}). 

Both ion signals reveal clear differences in their line shape and sign of amplitudes, which points to different excitation pathways leading to the ionic products. In accordance with their hypothesis, a laser intensity analysis shows different high-order multiphoton processes for the detected cationic signals (8'th order for NO$_2^+$ and 10'th order for NO$^+$). The high-order nonlinearity reveals the challenging experimental conditions in this study, however, also leads to ambiguities due to many overlapping high-order pathways that cannot be discriminated. This might be resolved in the future by incorporating additional phase cycling steps or extended pulse sequences. 

In general, the multiphoton ion-mass-detected 2D spectroscopy approach shows the potential to study ionization pathways and ultrafast autoionization processes in highly excited Rydberg manifolds of gaseous molecules, where 2D spectroscopy may provide additional information about transient intermediate states and improves the analysis of complex high-order multiphoton processes. 

\begin{figure}
\centering
  \includegraphics[width=0.8\linewidth]{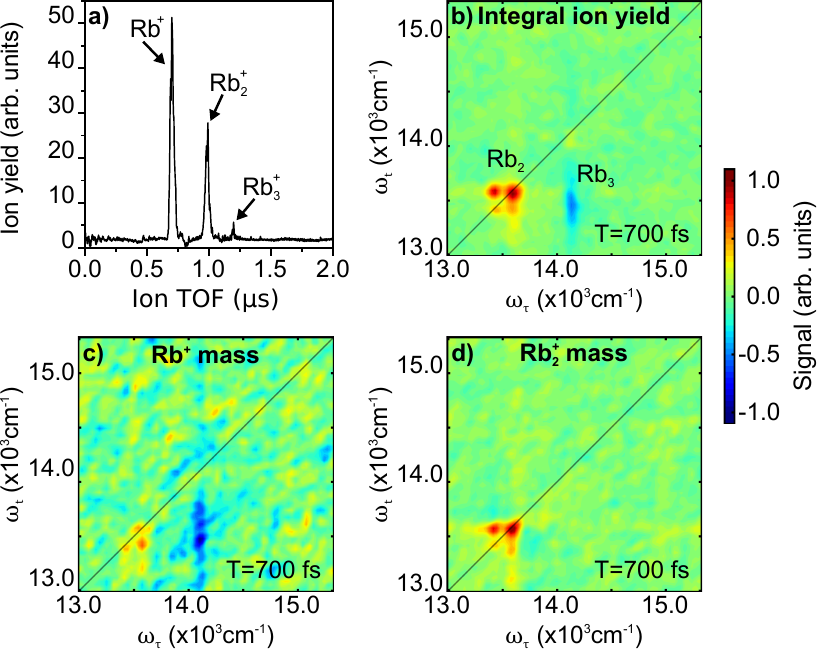}
  \caption{Mass-resolved 2DES. (a) Ion time-of-flight (TOF) trace of photoionized rubidium molecules desorbed from helium nanodroplets (recorded with additional, delayed ionization laser). (b) 2D spectrum recorded without mass selection. (c-d) 2D spectrum recorded at masses Rb$^{+}$ and Rb$_2^{+}$, respectively. 
   }
\label{fig:IonTOF}
\end{figure}

In the same fashion to this study, we have combined ion-mass detection with the phase modulation approach. 
In an early demonstration, we combined phase-modulated quantum interference measurements with a quadrupole ion mass filter for sensitive detection of RbHe excimer formations\,\cite{bruder_phase-modulated_2015}. For our 2DES experiments, we used instead an ion-TOF spectrometry arrangement (Fig.\,\ref{fig:IonTOF}). The 2D spectra of specific masses are recorded by means of TOF-gating using boxcar integrators. To this end, boxcar windows are placed on the respective mass peaks in the ion-TOF transients and the boxcar output is fed into the lock-in detection and processed as discussed in section 3. 

For the above discussed study of Rb molecules, the ion-TOF distribution (Fig.\,\ref{fig:IonTOF}a) shows three significant peaks, which correspond to the Rb$^{+}$, Rb$_2^{+}$ and Rb$_3^{+}$ species. Examples of obtained 2D spectra (for population time $T=700$\,fs) are shown in (Fig.\,\ref{fig:IonTOF}c,d) along with a 2D spectrum recorded without mass-gating as a reference (Fig.\,\ref{fig:IonTOF}b). From these measurements, we can learn more details about the photo-induced dissociation dynamics and can identify specific dissociation channels. 

For the used laser frequencies, Rb atoms may only be ionized via off-resonant three-photon excitation, which is a negligible process for the applied low laser intensities. Thus, the Rb$^+$ cations reflect the dissociation products of Rb$_2$ and Rb$_3$ molecules and carry the nonlinear response of both parent molecules, respectively, as reflected in the 2D spectrum of Fig.\,\ref{fig:IonTOF}c. Here, we directly see, that the ESA pathway in the Rb$_3$ molecule leads to a dominant production of Rb$^+$ ions, whereas neither dimer (Fig.\,\ref{fig:IonTOF}d) nor trimer ions (not shown) are detectable for this excitation/ionization channel. This stands in contrast to the photodynamics in the Rb$_2$ molecule, where excitation to the $(1)^3\Pi_g$ manifold and subsequent ionization leads only to a small production of Rb$^+$ fragments and most signal is detected in the Rb$_2^+$ 2D spectrum. 

Interestingly, the Rb$_2$ ESA$_1$ excitation pathway is absent in all ion-detected measurements but can be clearly observed in photoelectron measurements (Fig.\,\ref{fig:Rb2Rb32DES}d). This may be explained by a direct transition into the ion continuum or an ultrafast autoionization induced via the ESA pathway. Both processes would lead to immediate ionization by pulses 3, 4 before desorption of the excited molecule from the droplet surface has taken place. This is followed by the solvation of the cation into the helium droplet accompanied by the formation of a surrounding high density shell of He atoms (\textit{snowball} formation), which can be detected at large masses ($> 500$\,amu)\,\cite{vangerow_dynamics_2015}. In contrast, the lower-lying ESA$_2$ transition can be observed in ion-detected spectra at certain population times (Fig.\,\ref{fig:Rb2Rb32DES}c), which indicates the absence of a coupling to the ion continuum for slightly lower-lying states. 

Eventually, the sum of the 2D spectra recorded for individual mass-species (Fig.\,\ref{fig:IonTOF}c,d) matches with the spectrum obtained from integral ion yields (Fig.\,\ref{fig:IonTOF}b), confirming that no signals are omitted in the mass-selective detection. 

Both examples, from the Brixner group and our group, show the added information, one may gain by combining coherent 2D spectroscopy with mass spectrometry. As an advantage over conventional pump-probe mass spectrometry, 2DES provides spectral information for pump and probe steps and is able to track the coherent molecular dynamics. This information may help to decipher complex ultrafast photoreactions including involved dissociative dynamics.

\subsection{State-selectivity and modulation contrast}

\begin{figure}
\centering
  \includegraphics[width=0.8\linewidth]{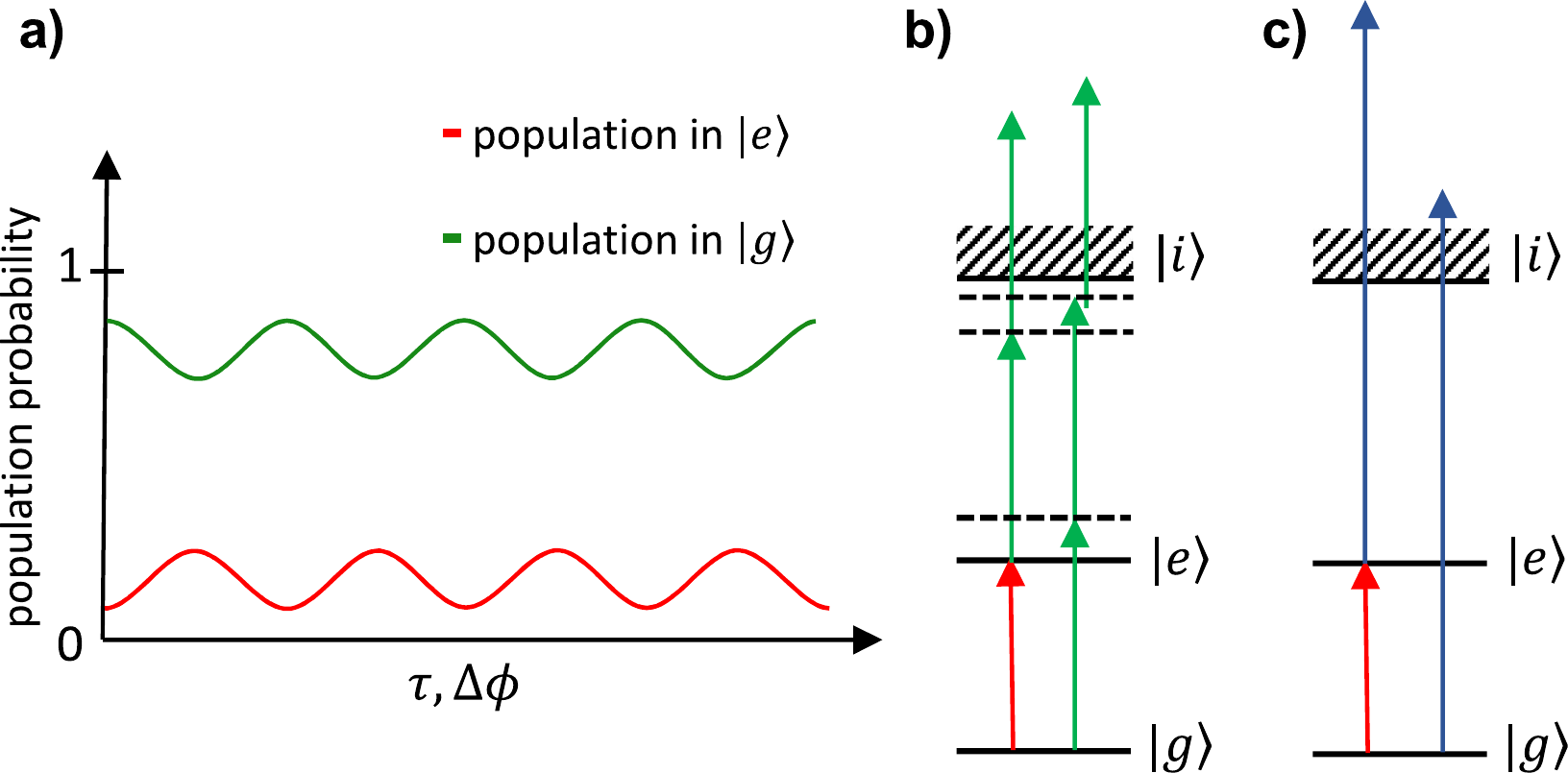}\vspace*{0mm}
  \caption{Comparison of photoionization schemes and modulation contrast. a) Oscillation of the population probability in a two-level system as a function of the relative delay and phase of excitation pulses. b,c) Mapping of populations to the ion continuum by photoionization. A nonlinear population state is induced by the 4-pulse sequence used in action-detected 2D spectroscopy (red arrow), which is detected via multiphoton ionization (a) or one-photon ionization (b). The ionization of ground ($\ket{g}$) and excited states ($\ket{e}$) is shown by green/blue arrows. Dashed lines indicate resonant or virtual intermediate states in the multiphoton ionization. 
  }
\label{fig:Ionization}
\end{figure}

A crucial point in all action-detected 2DES is the modulation contrast. 
The measurements rely on the detection of small modulations of a nonlinear population, induced by the coherent interactions of four optical pulses. Systematic modulation of the pulses' phase induces an alternation of the probability to reach the final population state. By applying well-defined phase patterns on the excitation pulses, the desired third-order nonlinear response of the system can be isolated from population signals modulated at different patterns and non-modulated background contributions. 

The sensitivity of this detection concept critically depends on the detection contrast between the final excited population state and the complementary state (e.g. the ground state). This becomes clear when considering the simple case of a two-level system, where a coherence induced between both states leads to a complementary (antiphase) oscillation of the excited and ground state population as a function of the relative pulse delay/phase (Fig.\,\ref{fig:Ionization}a). If excited and ground state populations are detected with equal probability, the modulation contrast is lost and a constant signal is measured. Hence, a high contrast between the detection efficiency of both states is important. 

An alternative explanation is given by the Feynman diagrams in Fig\,\ref{fig:pathways}b. For each process, always two complementary pathways exist (denoted as 1 and 2), which differ in the final population state (being $\ket{e}$ and$\ket{g}$ or $\ket{f}$ and$\ket{e}$). Both pathways are identical except for the 4'th interaction, leading to an antiphase modulation of the signal yield. If both contributions (i.e. final population states) are detected with the same efficiency, the modulations cancel each other leading to a depletion of the signal. 

In case of multiphoton ionization of the sample, a reasonably strong modulation contrast is naturally given, since complementary states require different numbers of photons for the ionization (Fig.\,\ref{fig:Ionization}b). Here, care must be taken to avoid saturation and hence loss of modulation contrast due to an intense ionization laser which is close to resonance with the probed transition ($\ket{g}\rightarrow \ket{e}$). The least photons are required for the ionization of ESA pathways ending in a high-lying population state (ESA2 in Fig\,\ref{fig:pathways}b), due to which these signals are generally amplified in mulitphoton detection. Likewise, any state may be selectively amplified by suitable choice of ionization laser wavelength and consideration of resonant intermediate levels. This property might be exploited to amplify and discriminate certain features in the 2D spectra in order to further disentangle overlapping spectral features from different species (Fig.\,\ref{fig:Rb2Rb32DES}). 

While multiphoton ionization may add complexity to the measurements due to the influence of intermediate levels, direct one-photon ionization has the advantage of mapping all populations directly to the ion continuum (Fig.\,\ref{fig:Ionization}c). Yet, one-photon ionization with UV pulses tends to generate high background signals by ionizing constituents of background gas and ground state molecules. Moreover, the ionization probability of a one-photon process solely depends on the bound-free wavefunction overlap which generally shows only small variations between different bound states and thus causes a loss of modulation contrast in the detection. This problem can be solved with photoelectron spectrometry, where modulation contrast is readily recovered by selecting the photoelectrons only from specific populations (and excluding their complementary population states). We have recently implemented such a scheme based on photoelectron TOF measurements in a magnetic bottle spectrometer. 
Our first results show an increased detection efficiency as compared to other methods\,\cite{bruder_delocalized_2019}.

\section{Outlook}
In conclusion, action-detected 2DES has already demonstrated a high degree of sensitivity in measurements including both high spectral resolution and full femtosecond dynamics, thereby facilitating the combination with extremely dilute samples like molecular beams at target densities down to $10^7$\,cm$^{-3}$ or even below. The combination with HENDI yields studies at millikelvin temperatures and unprecedented high resolution enabling a new level in the interpretation of dynamics in comparison with theory. The HENDI technique provides a plethora of tailored model systems ranging from weakly-bound van der Waals molecules, microsolvated systems up to specifically designed large organic complexes. Other prospective targets like size-selected free ions or charged clusters may open alternative avenues where 2DES could provide new insight, thus, being the ideal playground to study photoinduced ultrafast processes.

Furthermore, with the capabilities demonstrated already, coherent spectroscpoy may find its way into new disciplines. E.g., the field of  Quantum Optics could pivotally gain from corresponding new approaches. Ultracold ensembles in optical lattices or cold Rydberg gases might be studied with respect to the full dynamics of all coupled states, or Markovian vs.\ non-Markovian dissipation when probing interactions with external modes. On the other hand, high sensitive detection methods, as introduced above, are prerequisite for experiments on non-trivial quantum effects exploiting e.g.\ correlation experiments with single photon light sources. With the ongoing development of high repetition rate and high photon flux sources, exciting experiments have come into reach.

\subsection{New dimensions accessible in the gas phase}

The photoionization detection proved to be a valuable extension in 2DES. The variety of detection schemes available at UHV conditions grants a high control of the detection process, thereby simplifying 2D spectra through precise selectivity, and adding a considerable amount of complementary information. The incorporation of further detection schemes will open extra dimensions in multidimensional spectroscopy schemes. Velocity-map-imaging combined with sophisticated online data processing offers not only selected photoelectron energies but also electron emission angles as extra dimension. VMI ion images can provide directionality in dissociation processes. Alignment of molecules in strong laser fields or by means of well-established molecular beam methods gives access to the stereodynamics of chemical reactivity. Finally, detecting multiple electrons and ions in coincidence and correlation methods would even further enhance dimensions and selectivity in new types of studies.

However, most of the advanced techniques are limited to low data acquisition rates, which are still in the kHz range for digitizing time-of-flight traces or even below 50\,Hz for CCD-cameras. This conflicts with an ideal  high SN ratio at high repetition rates and shot-to-shot modulation. As a first step into this direction we have already shown that undersampling schemes allow to combine high modulation frequencies with low sampling rates\,\cite{Bruder:18}. Another issue is to properly process the full amount of information (e.g. multiple peaks in TOF) without an excessive number of lock-in hardware. In this direction software based lock-in algorithms\,\cite{Karki_lockin_2013} as well as advanced post-processing methods are on the way to readily scale the number of demodulators and to enable detailed shot-to-shot data processing.

\subsection{Coherent spectroscopy methods in the extreme-UV spectral range} 
The introduced phase-cycling methods inherently enable the detection of higher harmonic processes, e.g.\ in high-harmonic demodulation at phase-modulation experiments\,\cite{bruder_efficient_2015}. Apart from studying multiple quantum coherences (MQC)\cite{bruder_efficient_2015, bruder_delocalized_2019, Yu_2019}, it has been demonstrated, that this can be employed for phase-modulation experiments using light generated in higher harmonic generation (HHG) processes\,\cite{bruder_phase-modulated_2017}. In a recent approach, we successfully performed an extension of this work at the seeded Free-Electron Laser (FEL) FERMI, where by means of phase modulation of the UV seed laser, attosecond wave packet interferometry at XUV photon energies (28\,eV) was done\,\cite{wituschek_tracking_2019}. Control of femtosecond  pulse timing and CEP at higher harmonics in the XUV has been demonstrated, only acting on the fundamental UV laser pulses. In view of the rich options based on HHG XUV light sources covering  pulse durations down to the attosecond range, the prospective extension of multidimensional coherent methods at high photon energies would open a new field including inner shell processes, site specifity in molecular complexes, and attosecond time resolution.

Of course, many specific aspects realizing these kind of new experiments are still to be worked out and will remain challenging. However, the recent steps in 2DES are encouraging for many more exciting experiments that are underway. 

\ack
Funding by the European Research Council within the Advanced Grant "COCONIS" (694965), by the Bundesministerium f{\"u}r Bildung und Forschung (05K16VFB) and by the Deutsche Forschungsgemeinschaft (IRTG 2079) are acknowledged.

\section*{References}
\bibliographystyle{CPprsty}
\bibliography{CMDSgasphase_1,LB_added_citations}

\end{document}